\documentclass[aps,pre,preprint,floatfix]{revtex4}
\usepackage{dcolumn}%
\usepackage{graphicx}
\usepackage{amsmath}
\usepackage{amsfonts}
\usepackage{amssymb}
\usepackage{graphicx}
\usepackage{bm}%
\usepackage{verbatim}

\newcommand{\bq}{\begin{equation}}
\newcommand{\eq}{\end{equation}}

\begin{document}
\title{Heat pulse propagation in chaotic 3-dimensional magnetic fields}
\author{Diego del-Castillo-Negrete}
\affiliation{Oak Ridge National Laboratory \\ Oak Ridge, Tennessee 37831-8071, USA}
\author{Daniel Blazevski}
\affiliation{Institute for Mechanical Systems \\
ETH Zurich, 8092 Zurich, Switzerland}
\begin{abstract}
Heat pulse propagation in $3$-dimensional chaotic magnetic fields is studied by  solving numerically the parallel heat transport equation using a Lagrangian-Green's  function (LG) method. The LG method provides an efficient and  accurate technique that circumvents known limitations of  finite elements and finite difference methods. The main two problems addressed are: (i) The dependence of the radial transport of heat pulses on the level of magnetic field stochasticity (controlled by the amplitude of the magnetic field perturbation, $\epsilon$); and (ii) The role of reversed shear magnetic field configurations on heat pulse propagation. 
In all the  cases considered there are no magnetic flux surfaces. However, the radial transport of heat pulse is observed to depend strongly on $\epsilon$ due to the  presence of high-order magnetic islands and Cantori. These structures act as quasi-transport barriers which can actually preclude the radial penetration of heat pulses within physically relevant time scale.  The dependence of the magnetic field connection 
length, $\ell_B$, on $\epsilon$ is studied in detail.  
Regions where $\ell_B$ is large, correlate with regions where the  radial propagation of the heat pulse slows down or stops. 
The decay rate of the temperature maximum,  $\langle T \rangle_{max}(t)$, the time delay of the temperature response as function of the radius, $\tau$, and the 
radial heat flux $\langle {{\bf q}\cdot {\hat e}_\psi} \rangle$,  are also studied as functions  of the magnetic field stochasticity and $\ell_B$. In all cases it is observed that  the scaling of $\langle T \rangle_{max}$ with $t$ transitions from sub-diffusive, $\langle T \rangle_{max} \sim t^{-1/4}$, at  short times ($\chi_\parallel  t< 10^5$) to a significantly slower, almost flat scaling at longer times ($\chi_\parallel  t > 10^5$). A strong dependence on $\epsilon$ is also observed on $\tau$ and $\langle {{\bf q}\cdot {\hat e}_\psi} \rangle$. Even in the case when there are no flux surfaces nor magnetic field islands, reversed shear magnetic field configurations exhibit unique transport properties. 
The radial propagation of heat pulses in fully chaotic fields considerably slows down in the shear reversal region and, as a result, the delay time of the temperature response in reversed shear configurations 
is about an order of magnitude longer than the one observed in monotonic $q$-profiles. 
The role of separatrix reconnection of resonant modes in the shear reversal region, and the role of shearless Cantori in the observed phenomena are also discussed. 
\end{abstract}
\maketitle

\section{Introduction}
\label{introduction}

The use of magnetic fields to confine high temperature plasmas is considered the most promising mechanism for achieving controlled nuclear fusion.  
Ideally, magnetic field configuration should consist of nested flux surfaces. However, in practice, external coils or dynamical processes in the plasma can create magnetic perturbations that lead to chaotic magnetic fields and the eventual destruction of flux surfaces.  In some cases these perturbations can  be beneficial. For example, resonant magnetic perturbations are regularly used to control and suppress edge-localized modes (ELMs) \cite{evans_2008}. 
The interplay between magnetic field chaos and transport is also critical 
in the  control of heat fluxes  at the divertor \cite{Schmitz_2008}. 
Because of this, it is of significant interest to understand at a quantitative level the effect of magnetic field stochasticity on heat transport. 
Our goal in this paper is to address this problem in the context of radial propagation of heat pulses in weakly chaotic and fully chaotic magnetic fields with monotonic and reversed shear configurations. 


Heat transport in magnetized plasmas typically exhibits a strong anisotropy
due to the disparity between the parallel (i.e., along the magnetic  field) conductivity, $\chi_\parallel$, and the perpendicular conductivity, $\chi_\perp$. In particular, in fusion plasmas, the ratio, $\chi_\parallel/\chi_\perp$,  might exceed $10^{10}$.  Motivated by the need to understand transport in this extreme anisotropic regime, we focus on the study of purely parallel transport, i.e. we assume $\chi_\perp=0$.

It is well-recognized that this extreme anisotropic regime presents highly nontrivial numerical challenges \cite{DL}. In particular, when the magnetic field is not aligned to the numerical mesh (which is the generic case, specially in the study of $3$-dimensional chaotic magnetic fields) discrete spatial representations of the parallel transport equation typically suffer from numerical pollution. Another  issue is the  development of nonphysical negative temperatures from  positive definite initial conditions due to the accumulation of numerical errors. A related limitation faced by grid-based methods in the 
extreme anisotropy regime is the unavoidable eventual lack of spatial resolution due to filamentation. This is  a particularly pervasive issue in the case of chaotic magnetic fields in which, by definition, nearby point in the temperature initial condition field separate exponentially due to parallel transport at the rate dictated by the corresponding Lyapunov exponent. 


To circumvent these problems, Refs.~\cite{DL,DL_pop} proposed a Lagrangian-Green's function (LG) method as an 
alternative to the widely used, but limited, Eulerian grid-based approaches
for the solution of the anisotropic heat transport equation.
The numerical study presented here is based on 
this method. The LG method, which will be briefly review in Sec.~\ref{LG_review}, is
particularly well-suited to the problem of transport in highly chaotic  $3$-dimensional magnetic fields which play a central role in the present study. 
In particular, the LG method provides an accurate and efficient algorithm that,
by construction, preserves the positivity of the temperature evolution, avoids completely the pollution issues encountered in grid-based methods, and can cope with the temperature filamentation at any scale. 
Another advantage of the LG method is the relative ease with which different parallel heat flux closure models can be incorporated. In the strong collisional limit, parallel transport is diffusive and  the heat flux closure is the well-known Fourier-type prescription according to which the flux depends on the local properties of the temperature gradient. However, in the case of low collisionality plasmas, the closure is non-local and the parallel transport equation involves an integro-differential operator which is expensive to evaluate using Eulerian methods. In the LG method the difference between local and non-local closures only reflects in the specific functional form of the Green's function. Because of this (except for convergence issues related to the specific  asymptotic behavior of the Green's function) in the LG method the computation of the temperature evolution is algorithmically the same for any type of closure for which the Green's function is known analytically or numerically. 


The main goal of this paper is to study the role of  magnetic field stochasticty  on the radial propagation of localized heat pulses. 
We limit attention to parallel transport and compute the temperature evolution using the LG method assuming a diffusive, local parallel flux closure.  
In addition to the level of stochasticity, we study the dependence of the 
radial propagation of heat pulses on the $q$-profile. Of particular interest is to compare transport in monotonic $q$-profiles with  transport in reversed shear configurations in which the $q$-profiles is not monotonic. 
Reversed shear configurations are interesting from a practical perspective because they typically  exhibit very robust transport  barriers in toroidal plasma confinement devices \cite{italians_1,italians_2}. 
At a more fundamental level, the study of transport and magnetic field sotochasticity in reversed shear configurations  is interesting 
because the destruction of magnetic flux surfaces in the regions where the magnetic shear vanishes is fundamentally different to what happens in regions where the magnetic shear is finite. This result,  which goes beyond plasma physics application, was originally discussed in the general context of Hamiltonian chaos theory and fluid mechanics in Refs.\cite{Dnt1,dcn_1993} where  the resilience of shearless 
Kolmogorov-Arnold-Moser (KAM) curves was numerically found and the transition to chaos was shown to belong to a new universality class.
Previous application of this generic Hamiltonian dynamics result to magnetically confined plasmas include the early works in Refs.~\cite{Dnt0,balescu,dcn_2000}, and the later studies reported in 
Refs.~\cite{firpo_2011,ibere_2011}.  More recently,  in 
Ref~\cite{dan_diego_2013} 
a study, also based on the LG method, was presented addressing the 
role of separatrix reconnection and the resilience of shearless barriers (two key signatures of reversed shear magnetic field configurations) on heat transport. 
The study presented here focusses on following the spatio-temporal evolution of localized  heat pulses introduced in the edge of the computational domain. 
In this sense, the  setting of the numerical simulations resembles the rationale adopted in experimental perturbative transport studies
where the evolution of the plasma temperature response to edge localized pulses is measured.  For example, in Ref.~\cite{Ida_2013}, the changes in the radial propagation speed of heat pulses were used as a diagnostic to characterize the level of the magnetic field  stochasticty in the Large Helical Device (LHD) experiment.  Going beyond previous works, here we present 
the first study on the detailed dependence of the radial propagation of heat pulses on the level of stochasticity and on the $q$-profile based  on the direct numerical solution of the time dependent parallel heat transport equation in 3-dimensional magnetic fields.

The rest of this article is organized as follows. The next section contains a brief  review of the Lagrangian Green's function (LG) method.  
Section~\ref{models} presents the magnetic field models used in the solution of the parallel heat transport equation. The core of the  numerical results on heat pulse propagation in monotonic $q$-profile and reversed shear chaotic field  are presented in Sec.~\ref{results}. Section~\ref{conclusions} contains the conclusions. 

\section{Lagrangian-Green's function (LG) method}
\label{LG_review}
 
In this section we review the method to solve the parallel heat transport equation developed in Refs.~\cite{DL,DL_pop}.
The starting point is the heat transport equation,
\begin{equation}
\label{eq_II_1}
 \partial_t T = - \nabla \cdot \mathbf{q} + S \, ,
\end{equation}
where $S$ is a source, and $\mathbf{q}$ is the heat flux consisting of 
a parallel (along the magnetic field) and a perpendicular 
component, ${\bf q}= q_\parallel {\bf \hat b} + {\bf q}_\perp$, where ${\bf \hat b}={\bf B}/|B|$ is the unit magnetic field vector. 
We will limit  attention to parallel heat transport in the extreme
anisotropic regime 
i.e., we will assume ${\bf q}_\perp=0$. This assumption is motivated 
by the strong anisotropy typically encountered in magnetized
 fusion plasmas in which the ratio $\chi_\parallel/\chi_\perp$ can be of the order of  $10^{10}$,  where $\chi_\parallel$ and $\chi_\perp$ denote the parallel and perpendicular conductivies.
Writing the parallel heat flux as 
\bq
\label{eq_II_2}
q_\parallel= \chi_\parallel {\cal Q} \left[ T\right] \, ,
\eq
where ${\cal Q}$ denotes a general differential or integro-differential, possibly nonlinear, operator, Eq.~(\ref{eq_II_1}) becomes
\bq
\label{eq_II_3}
\partial_t T + \chi_\parallel \left( \nabla \cdot {\bf \hat b}\right) {\cal Q} = -\chi_\parallel \partial_s {\cal Q}  + S\, ,
\eq
where $ \nabla \cdot {\bf \hat b}=-\left(\partial_s B \right)/B$,
$\partial_s=\hat{ \bf b} \cdot \nabla$ denotes the directional derivative along the magnetic field line with $s$ the arc-length parameter, and  we have assumed that the parallel diffusivity is constant along the field line, i.e., 
$\partial_s \chi_{\parallel}=0$. 

\begin{figure}
\includegraphics[width=0.75 \columnwidth]{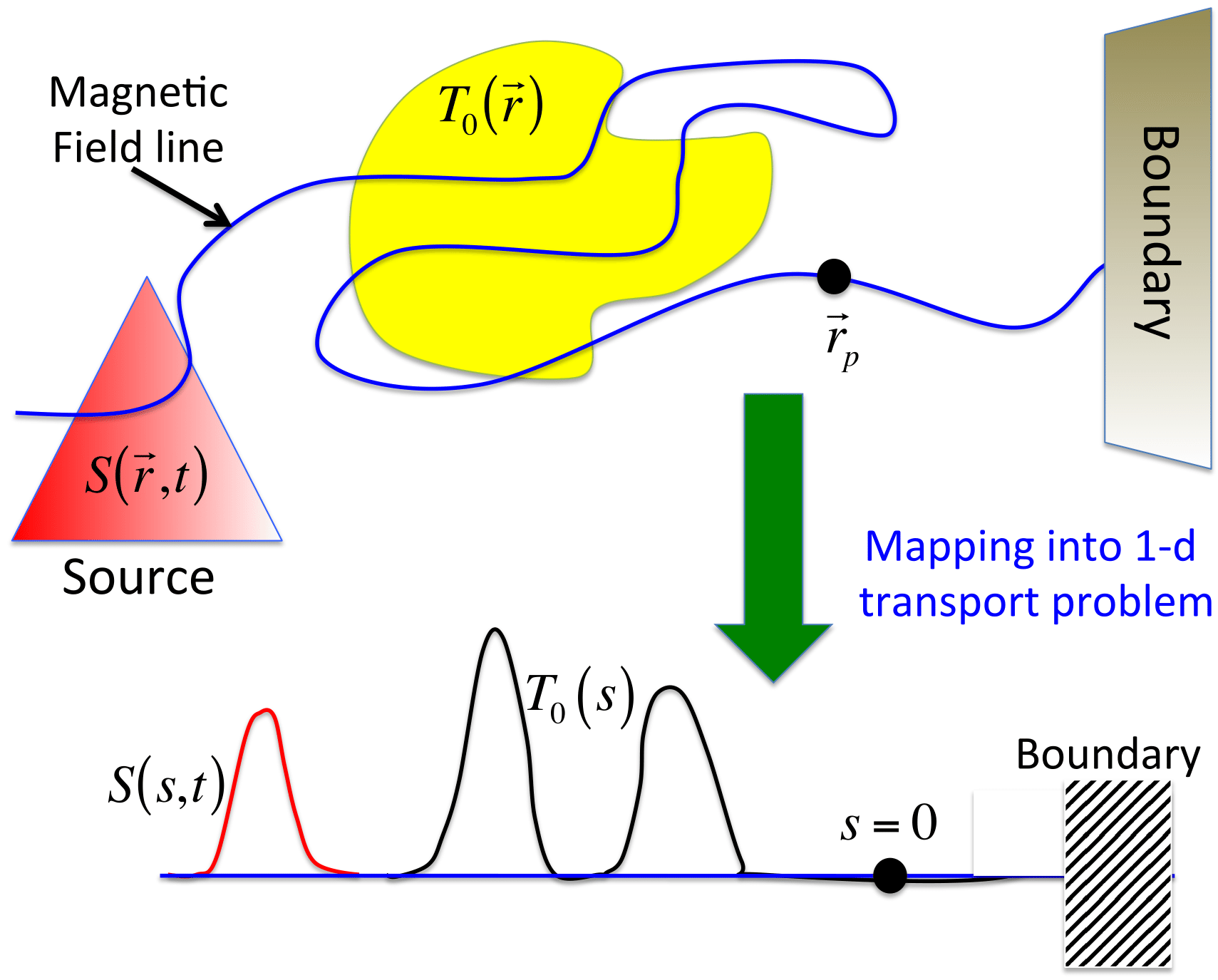}
\caption{(Color online) Lagrangian-Green's function (LG) method for anisotropic heat transport in general magnetic fields \cite{DL,DL_pop}.  
In the absence of perpendicular transport,  ${\bf q}_\perp=0$, the temperature at  point ${\bf r}_0$, at time $t$, depends only on the heat transported along the unique magnetic field line passing through ${\bf r}_0$. 
If  ${\cal G}_\alpha$ is the Green's function of the parallel heat transport operator, $T({\bf r}_0,t)$ is computed directly by evaluating the integrals in
Eq.~(\ref{eq_II_10}) where ${\bf r}={\bf r}(s)$ is the magnetic field line trajectory parametrized by the arc-length, with ${\bf r}(s=0)={\bf r}_0$, and  $S({\bf r}(s),t)$ is the source.
}
\label{fig_LG_cartoon}
\end{figure}

In what follows we will 
neglect the second term on the left hand side of 
Eq.~(\ref{eq_II_3}). That is, we assume
$\left| \left( \partial_s B\right)/B\right| \ll \left| \left( \partial_s {\cal Q}\right)/{\cal Q}\right|$, and write the parallel transport equation as
\bq
\label{eq_II_4}
\partial_t T = -\chi_\parallel \partial_s {\cal Q}  + S\, .
\eq
This toroidal ordering approximation is commonly used in the study of magnetically confined plasmas in the presence of a strong guiding field. In particular,
it is a good approximation in cylindrical geometry with
${\bf B}={\bf B}_0+\epsilon {\bf B}_1$, where ${\bf B}_0$ is a helical field, for which $ \partial_s B_0=0$, and the 
perturbation satisfies, $\epsilon B_1\ll B_0 $, which is the case of interest in the present paper.  

For high collisionality plasmas, parallel transport is typically dominated by diffusion and a Fourier-Fick's type local flux-gradient relation of the form
\bq
\label{eq_II_5}
{\cal Q}\left[ T \right]=-\partial_s T\, ,
\eq
is assumed. Substituting Eq.~(\ref{eq_II_5}) into Eq.~(\ref{eq_II_4}), leads to the standard parallel  diffusion equation,
\begin{equation}
\label{eq_II_6}
 \partial_t T =  \chi_\parallel \partial^2_s T + S \, ,
\end{equation}
for collisional transport along magnetic field lines. 
On the other hand, in the case of  low collisionality plasmas, the parallel flux closure is in general nonlocal \cite{held_etal_2001}. That is, the heat flux at a point depends on the temperature distribution along the whole magnetic field line. 
An example is the non-local closure
\bq
\label{frac_flux}
{\cal Q}[T]=-\frac{\lambda_\alpha}{\pi} \int_0^\infty 
\frac{T\left(s+z\right)-T\left(s-z\right)}{z^\alpha} dz \, ,
\eq
where $\lambda_\alpha=-\pi (\alpha-1)/\left[ 2 \Gamma(2-\alpha) \cos (\alpha \pi/2)\right]$, and $1\leq \alpha<2$.
For $\alpha=1$, $\lambda=1$,  Eq.~(\ref{frac_flux}) reduces to the free streaming closure 
\cite{held_etal_2001}. 
In the more general case, $1\leq \alpha<2$, Eq.~(\ref{frac_flux}) describes super-diffusive transport along the magnetic field lines  governed by the parallel fractional diffusion equation
\bq
\label{fractional_eq}
\partial_t T = \chi_\parallel \partial^\alpha_{|s|} T \, ,
\eq
resulting from substituting Eq.~(\ref{frac_flux}) into Eq.~(\ref{eq_II_4})
\cite{del_castillo_2006}. 
The operator $\partial^\alpha_{|s|}$ denotes the symmetric fractional derivative of order $\alpha$ defined in Fourier space by the relation,
\bq
{\cal F}\left[ \partial_{|s|}^\alpha T \right ]=-|k|^\alpha  \hat{T} \, ,
\eq
where
${\cal F}\left[ T \right ]=\hat{T}(k)=\int_{-\infty}^{\infty} e^{i k x} T(x) dx$  denotes the Fourier transform.

The basic idea behind of the Lagrangian-Green's function (LG) method 
for the solution of the anisotropic heat transport equation for a time-independent magnetic field in the limit ${\bf q_\perp}=0$ is illustrated in Fig.~\ref{fig_LG_cartoon}.
Given an initial temperature distribution $T_0({\bf r})=T({\bf r},t=0)$, 
and a source $S({\bf r},t)$, the temperature at a given point in space ${\bf r}_0$, at a time $t$, is obtained by summing all the contributions of the initial condition and the source along the magnetic field line path ${\bf r}={\bf r}(s)$. That is, 
\bq
\label{eq_II_10}
T({\bf r}_0,t) = \int_{-\infty}^\infty T_0 \left[ {\bf r}(s') \right] {\cal G}_\alpha(s',t) ds' +
 \int_0^t dt'\, \int_{-\infty}^\infty ds' S \left[ {\bf r}(s'),t' \right] {\cal G}_\alpha(s',t-t') \, ,
\eq
where ${\cal G}_\alpha$ is the Green's function of the parallel transport equation 
(\ref{eq_II_4}), 
and  the magnetic field line trajectory, ${\bf r}(s)$, is  obtained from the solution of the initial value problem 
\begin{equation}
\label{eq_II_11}
\frac{d \mathbf{r}}{d s} = \hat{\mathbf{b}}\, , \qquad \mathbf{r}(0) = \mathbf{r}_0
\end{equation}
where $s$ is the  arc-length.

The first step in the  implementation of the LG method is to find the Green's function of the heat transport equation (\ref{eq_II_4})   with the appropriate boundary conditions. The simplest example corresponds to diffusive ($\alpha=2$)  parallel transport in a unbounded domain. In this case, as it is well-known, ${\cal G}_2$ is the Green's function 
 of the diffusion Eq.~(\ref{eq_II_6})
\begin{equation}
\label{eq_II_12}
 {\cal G}_2(s, t) = \frac{1}{\sqrt{2} \pi} (\chi_\parallel t)^{-1/2} \exp{\left(-\frac{s^2}{4 \chi_\parallel t} \right)} \, .
 \end{equation}
For general $\alpha$, ${\cal G}_\alpha$ is given by the Green's function of the fractional diffusion Eq.~(\ref{fractional_eq}),
\begin{equation}
\label{eq_II_13}
 {\cal G}_\alpha = \frac{1}{ (\chi_\parallel t)^{\frac{1}{\alpha}}} L_{\alpha} 
 \left[ \frac{s}{ (\chi_\parallel t)^{\frac{1}{\alpha} }} \right]
\end{equation}
where 
\bq
\label{eq_II_14}
L_\alpha(\eta)=\frac{1}{2 \pi} \int_{-\infty}^\infty e^{-\left | k \right |^\alpha - i \eta k} dk \, ,
\eq 
is the symmetric $\alpha$-stable Levy distribution. The case $\alpha = 1$, which corresponds  to the commonly used nonlocal free streaming closure, has the analytically simple expression
\begin{equation}
\label{eq_II_15}
{\cal G} _1(s,t) =  \frac{(\chi_\parallel t)^{-1}}{\pi} \frac{1}{1 + (s/\chi_\parallel t)^2} \, .
\end{equation}
More details on fractional diffusion and its applications to nondiffusive transport in
plasmas can be found in Refs.~\cite{del_castillo_2006,del_castillo_2008} and references therein.  Throughout the present paper we will limit attention to the case of local parallel  diffusive closures. 

The implementation of the LG method
requires three elements: an ODE integrator for solving the field line trajectories in 
Eq.~(\ref{eq_II_11}), an
interpolation procedure of the function $T_0({\bf r})$ on the field line, and a 
numerical quadrature to evaluate the Green's function integrals in Eq.~(\ref{eq_II_10}). 
These elements are relatively straightforward to implement numerically,  making the LG  algorithm a versatile, efficient, and accurate  method for the computation of heat transport in magnetized plasmas. By construction, the LG method preserves the positivity of the temperature evolution, and avoids completely the
pollution issues encountered in finite difference and finite elements algorithms. Also, because of the parallel
nature of the Lagrangian calculation, the formulation naturally leads
to a massively parallel implementation. In particular, 
the computation of  $T$ at  ${\bf r}_0$ at time $t$
does not require the computation of $T$  in the neighborhood of ${\bf r}_0$ or the computation of $T$ at previous times, as  it is the case in finite different methods.  Further details on the method and the numerical implementation can be found in Refs.~\cite{DL,DL_pop}.

In this brief summary we have  limited attention to purely parallel transport. However,  in the recent paper in Ref.~\cite{luis_jcp_2013}, the LG method has been extended to include finite perpendicular transport, i.e. $\chi_\perp \neq 0$.  
The key idea behind this extension is to formally include the perpendicular transport channel as part of an effective ``source", $S^*=S+ \chi_\perp \nabla_\perp T$, in Eq.~(\ref{eq_II_10}), and use the formal LG solution in Eq.~(\ref{eq_II_10}) to transform the heat transport equation into an integro-differential equation. The numerical solution of the integro-differential equation is based on an asymptotic-preserving  semi-Lagrangian operator-splitting algorithm consisting of two steps. The first step is the Eulerian solution of the perpendicular transport equation, and the second step is the solution of the parallel transport equation with source using the LG method.

\section{Magnetic field models}
\label{models}

We assume a periodic
straight cylindrical domain with period $L=2 \pi R$, and use
cylindrical coordinates $(r, \theta,z)$. 
 The magnetic field is given by
\begin{equation}
 \mathbf{B}(r, \theta, z) = \mathbf{B}_0 (r) + \mathbf{B}_1(r, \theta, z) \, ,
\end{equation}
where ${\bf B}_0(r)$ is a helical
field of the form, 
\bq
\label{B_model}
{\bf B}_0=B_\theta\,  \hat{\bf e}_\theta + B_z \hat{\bf e}_z \, , 
\eq
with $B_z$ constant,
$B_\theta=B_\theta(r)$, and $\mathbf{B}_1(r, \theta, z)$ is a perturbation. 
In all the calculations we assume $R=5$ and $B_z=1$.
The shear of the helical magnetic  field, i.e. the dependence of the azimuthal rotation of the field as function of the radius, is determined by the $q$-profile, $q(r)=r B_z/(R B_\theta)$. 
We consider two cases: magnetic field configurations with monotonic $q$-profiles, and reversed shear magnetic fields in which the derivative of the 
$q$-profile vanishes at a finite radius. 
In both cases it is assumed that the  perturbation, ${\bf B}_1$, has no $z$-component and is given by
 \bq
 \label{perturbation_B}
{\bf B}_1=\nabla \times \left[ A_z(r,\theta,z) \hat{\bf e}_z \right] \, ,
\eq
where the magnetic potential consists of the superposition of normal modes of the form
\bq
\label{perturbation}
A_z=\sum_{m,n} A_{mn}(r)  \cos \left( m \theta - n z/R  + \zeta_{mn}  \right) \, ,
\eq
where $\zeta_{mn}$ are constant phases.

\subsection{Monotonic $q$-profile}

In the monotonic $q$-profile case we assume
\bq
\label{B_theta}
B_\theta(r)=\frac{B (r /\lambda)}{1+(r/\lambda)^2} \, ,
\eq
which implies
\bq
\label{q_monotonic}
q(r)=q_0\left( 1 - \frac{r^2}{\lambda^2} \right ) \, .
\eq
In terms of the flux variable,
\bq
\label{psi_def}
\psi=\frac{r^2}{2 R^2} \, ,
\eq 
$q$ is a linear function of $\psi$. 
The mode amplitude functions are given by 
\bq
\label{eq_21}
A_{mn}(r)= \epsilon a(r) \left( \frac{r}{r_*} \right)^m \exp \left[
  \left( \frac{r_*-r_0}{\sqrt{2} \sigma}\right)^2 -\left(
  \frac{r-r_0}{\sqrt{2} \sigma}\right)^2 \right] \, , 
\eq with 
\bq
r_*=\lambda \sqrt{\frac{m}{n} \left( \frac{B R}{B_z \lambda}\right)-1}
\, , \qquad r_0=r_*-\frac{m \sigma^2}{r_*} \, .  
\eq 
By construction, $r=r_*$ corresponds to the location of the $(m,n)$
resonance, i.e., $q(r=r_*)=r_*
B_z/(R B_\theta(r_*))=m/n$.  The value of $r_0$ is chosen to guarantee
that at the resonance, the perturbation reaches a maximum, 
$dA_{mn}/dr (r=r_*)=0$, given by $A_{mn}(r_*)=\epsilon$. The prefactor
$(r/r_*)^m$ is included to guarantee the regularity of the radial
eigenfunction near the origin, $r\sim 0$ for the given $m$. The function, 
\bq
\label{a_fcn}
a(r)=\frac{1}{2} \left [
1 - \tanh \left( 
\frac{r-1}{\ell}
\right )
\right ] \, ,
\eq
is introduced to guarantee
the vanishing of the perturbation for $r=1$.  
In all calculations $\epsilon\sim {\cal O}\left(10^{-4}\right)$,
which, consistent with the tokamak ordering approximation, implies that
the variation of the field amplitude along the field line is of the order of
 $\left | \partial_s B / B \right | \sim 10^{-4}$.

\subsection{Reversed shear configuration}

To study the role of magnetic field chaos on heat transport in reversed shear configurations, we use the magnetic field model in Ref.~\cite{dan_diego_2013}. 
In this case, the $q$-profile is given by
\begin{equation}
\label{q_profile}
q(r) = q_0 \left[1+\lambda^{2}\left(r-\frac{{1}}{\sqrt{{2}}}\right)^{2} \right] \, ,
\end{equation}
which is non-monotonic in $r$ with a minimum at $r=r_{sl}=\frac{{1}}{\sqrt{{2}}}$.  
This implies a reversal of the magnetic field shear  which is positive for 
$r< r_{sl}$,  negative for $r > r_{sl}$,  and vanishes 
at the shearless point $r=r_{sl}$.
 The perturbation  has the same structure as in the monotonic $q$-profile case. That is, like in Eq.~(\ref{perturbation_B}), ${\bf B}_1$, has no $z$-component, and it is determined in terms of the magnetic potential according to Eqs.~(\ref{perturbation}).

For each $(m,n)$  the function $A_{mn}(\psi)$ is peaked at the resonance(s), $r^*$, defined by the condition $q(r^*) = m/n$. 
Due to the non-monotonicity of the $q$-profile, it is possible to have two, one, or no resonances  depending 
on the value of $m/n$.  
In the case when there are two resonances for a given $(m,n)$, 
we take
\begin{equation}\label{A_mn}
A_{mn}(r)  = \epsilon a(r) \left[A_{mn,1} + A_{mn,2} \right] \, ,
\end{equation}
where 
\begin{equation}
\label{A_term}
A_{mn,i} = C_{mn, i}r^m 
\exp  \left[- \left( \frac{r - r_{0i} }{\sqrt{2} \sigma} \right)^2 \right] \, ,
\end{equation}
for $i = 1, 2$.
For the function $a(r)$, defined in in Eq.~(\ref{a_fcn}), we choose $\ell=0.05$. 
The constants,
\begin{equation}\begin{split} 
& r_{0i} = r_{i*} - \frac{m \sigma^2}{r_{i*}}\, , \\
& C_{mn, i}  = \left( \frac{1}{r_{*i}} \right)^m \exp \left[ \left( \frac{r_{i*} - r_{0i}}{\sqrt{2} \sigma} \right) \right] \, ,
\end{split}\end{equation}
for $i=1, 2$, are chosen so that 
$A_{mn, i}$ has a maximum with unit amplitude at the location of the resonance, $r=r_{i*}$.
In the case when there is only one resonance for a given $(m, n)$,  we take
\begin{equation}\label{A_mn_single}
A_{mn}(r)  = \epsilon a(r) A_{mn,1}
\end{equation}
where $A_{mn, 1}$ is given in Eq.~(\ref{A_term}) with $i=1$.  
Finally, in the case when there are no resonances for $(m,n)$,  $A_{mn} = 0$.  

\section{Heat pulse propagation in monotonic  $q$ and reversed shear chaotic magnetic fields}
\label{results}

In this section we present the numerical results on the  propagation of heat pulses in the magnetic field configurations discussed in the previous section. In all the calculations the initial condition for the temperature, $T_0(\psi)$, consisted of a narrow  heat pulse localized near the edge of the domain $R^2 \psi \in (0, 0.5)$, 
\begin{equation}
\label{ic_T}
T_0(\psi)= \exp \left[ -  \frac{R^4(\psi - \psi_0)^2}{\sigma_p^2} \right]  \, ,
\end{equation}
with $R^2\psi_0 = 0.325$ and $\sigma_p = 0.008$. 
The temperature field was computed at different times on a 3-dimensional grid consisting 
of 50 points in $R^2 \psi \in (0, 0.5)$ , 45 points in $\theta \in  (0, 2\pi)$, and 2 points in the $z \in (0, 2\pi R)$ with $R=5$. 
The computation  at each grid point was performed by solving the parallel heat transport equation along the field line going through the corresponding point using the LG method discussed in Sec.~II. 
The numerical code used in the computations was originally developed in Refs.~\cite{DL,DL_pop}. 
The main object of study is the dependence on $\psi$ and $t$ of the temperature field averaged on $\theta$ and $z$, denoted as $\langle T \rangle (\psi,t)$. 
Throughout the present paper we limit attention to the case of parallel diffusive closure in Eq.~(\ref{eq_II_5}). Independently of the strength or type of perturbation, we will refer to the cases when the unperturbed axisymmetric system has a $q$-profile of the form in Eq.~(\ref{q_monotonic}) as ``Monotonic" and when the $q$-profile is of the form in  Eq.~(\ref{q_profile}) as  ``Reversed Shear".  

\subsection{Weakly chaotic fields in monotonic $q$-profile}

\begin{figure}
\includegraphics[width=0.50 \columnwidth]{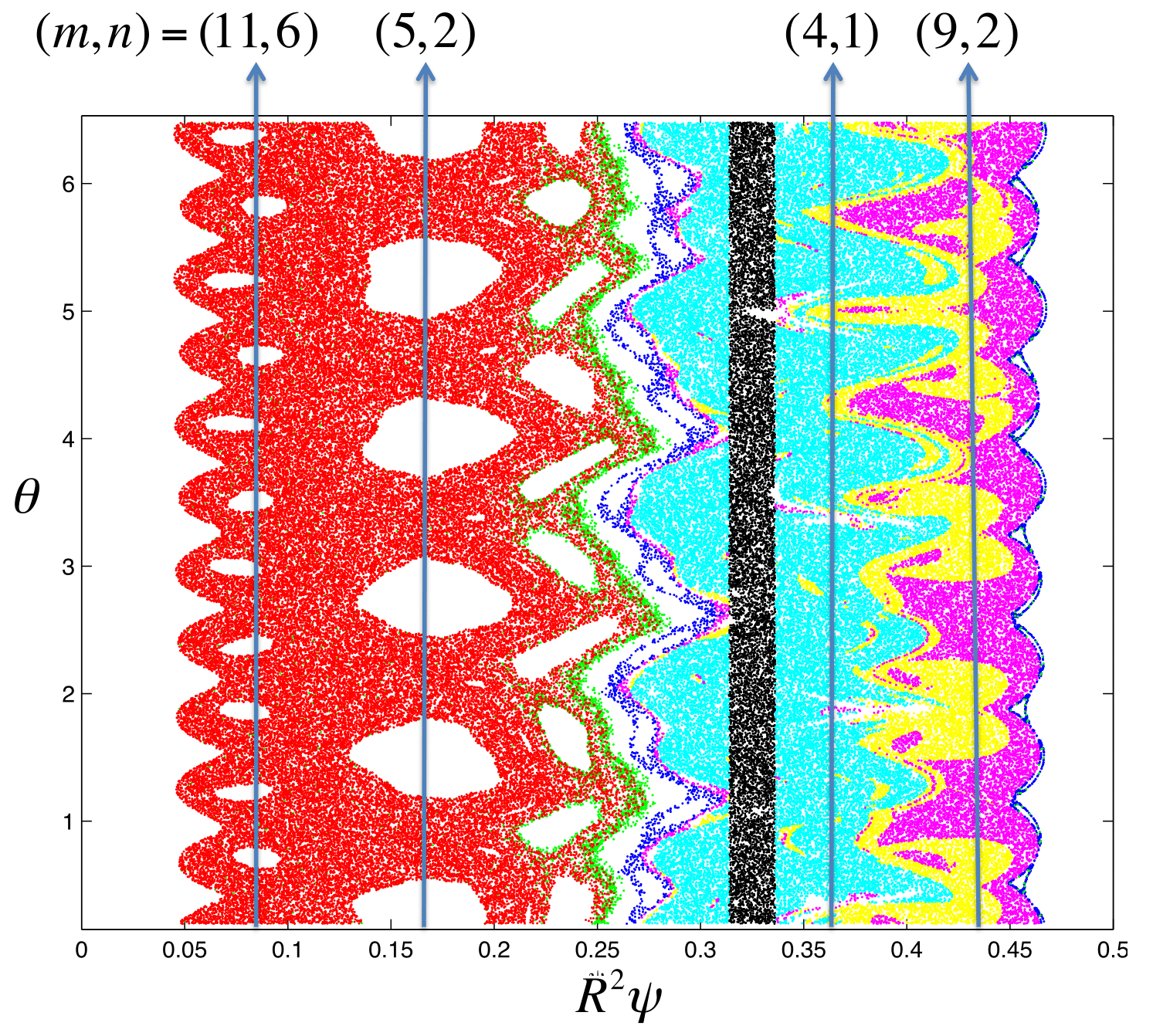}
\includegraphics[width=0.50 \columnwidth]{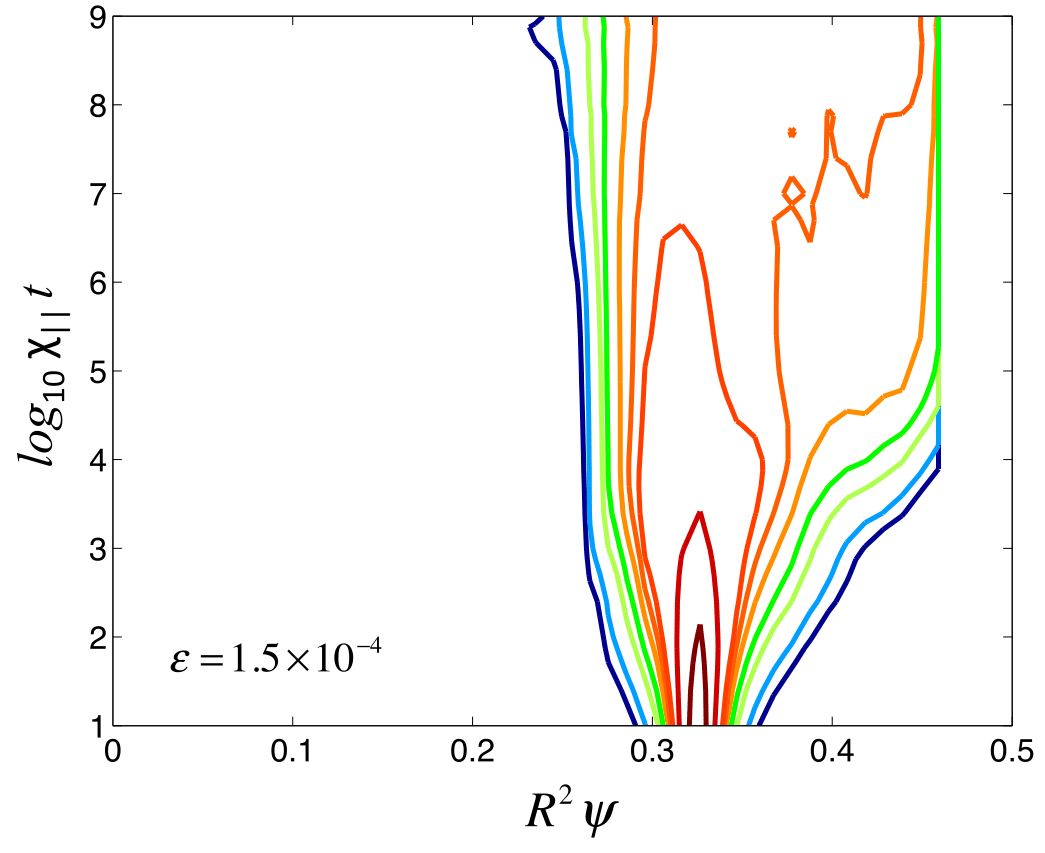}
\caption{
Magnetic field connection length and heat pulse propagation 
in a {\em weakly chaotic} magnetic field with {\em monotonic $q$-profile}. 
Top panel shows a Poincare plot  of the
magnetic field with monotonic $q$ profile in 
Eqs.~(\ref{q_monotonic}) perturbed by the four modes in Eq.~(\ref{four_modes})  
with amplitude $\epsilon=1.5 \times 10^{-4}$. The vertical arrows indicate the radial location of the resonances. The points are color coded by the value of magnetic field connection length, $\ell_B$, defined in Eq.(\ref{cl}). 
$\ell_B<5 $ (Cyan); $5<\ell_B<10 $ (Yellow); $10<\ell_B<10^2 $ (Magenta);
$10^2 <\ell_B< 10^3 $ (Blue); $10^3 < \ell_B<10^4 $ (Green);
$10^4 < \ell_B$ (Red). $\ell_B=0$ (Black) corresponds to the region where the pulse is introduced. 
The contour plot at the bottom panel shows the spatio-temporal evolution of the heat pulse. The contours correspond to 
(from red to blue) $T_0=\{ 0.5,\, 0.25,\, 0.10,\, 0.075,\,  0.05,\, 0.025,\, 10^{-2},\, 10^{-3},\,  10^{-4} \}$.}
\label{fig_CL_count_1_5}
\end{figure}

\begin{figure}
\includegraphics[width=0.50 \columnwidth]{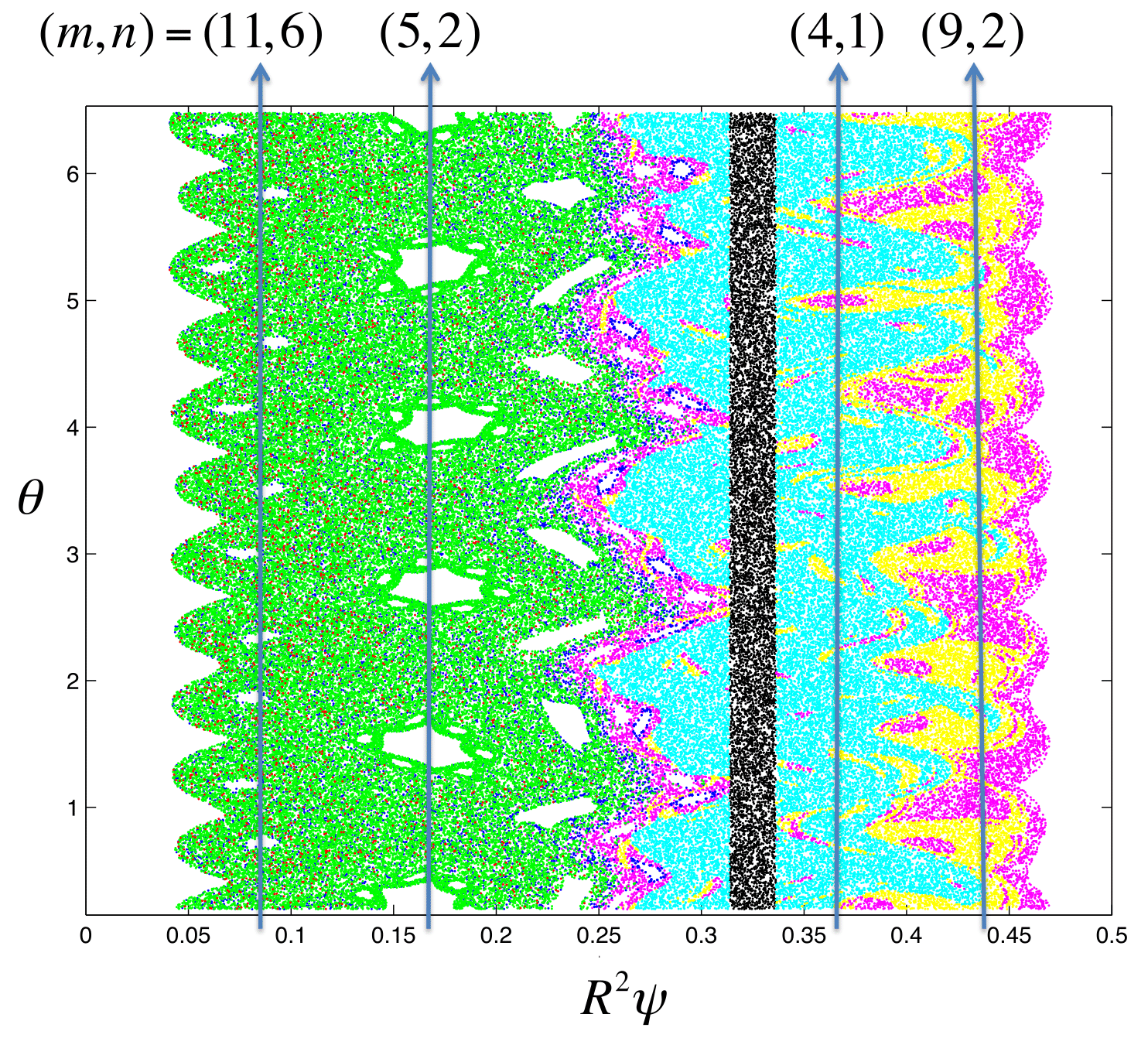}
\includegraphics[width=0.50 \columnwidth]{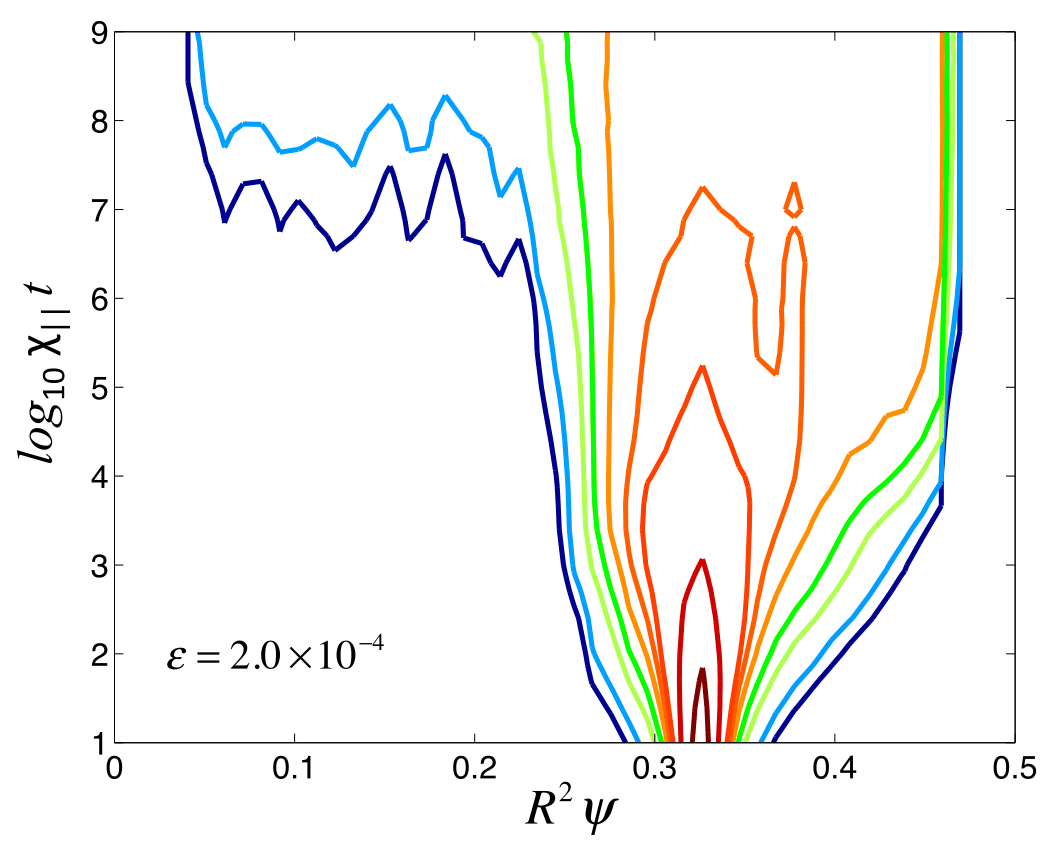}
\caption{Same as Fig.~\ref{fig_CL_count_1_5} but for $\epsilon=2 \times 10^{-4}$.}
\label{fig_CL_count_2_0}
\end{figure}

\begin{figure}
\includegraphics[width=0.50 \columnwidth]{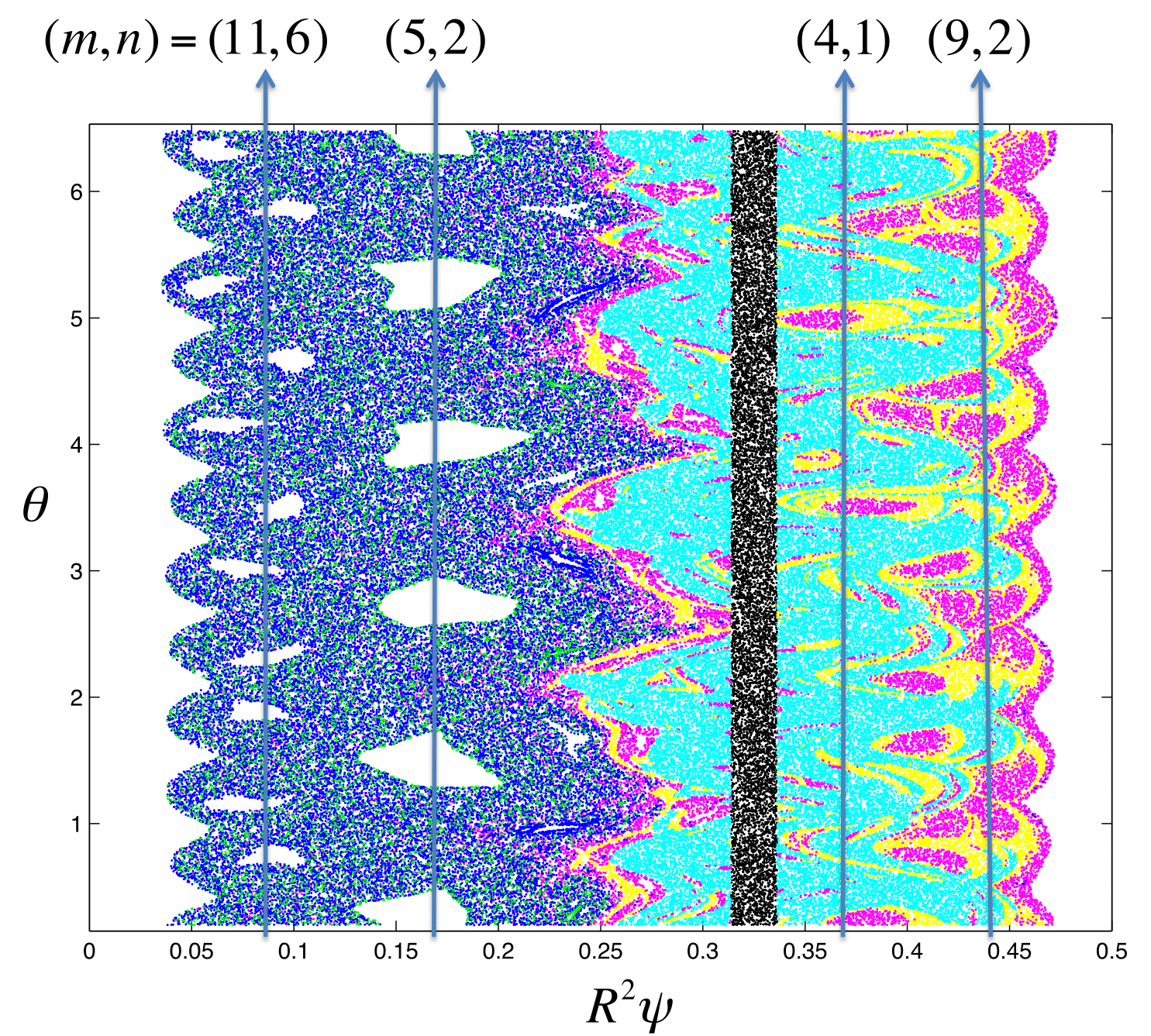}
\includegraphics[width=0.50 \columnwidth]{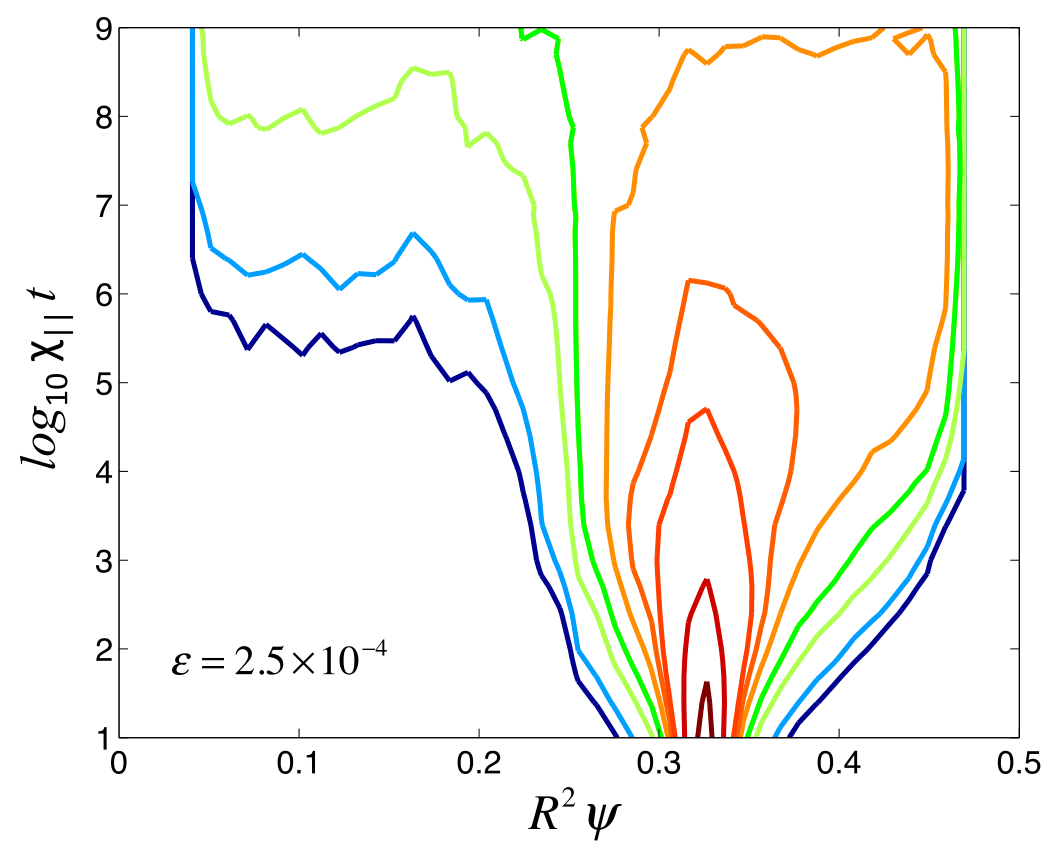}
\caption{Same as Fig.~\ref{fig_CL_count_1_5} but for $\epsilon=2.5 \times 10^{-4}$.}
\label{fig_CL_count_2_5}
\end{figure}

\begin{figure}
\includegraphics[width=0.50 \columnwidth]{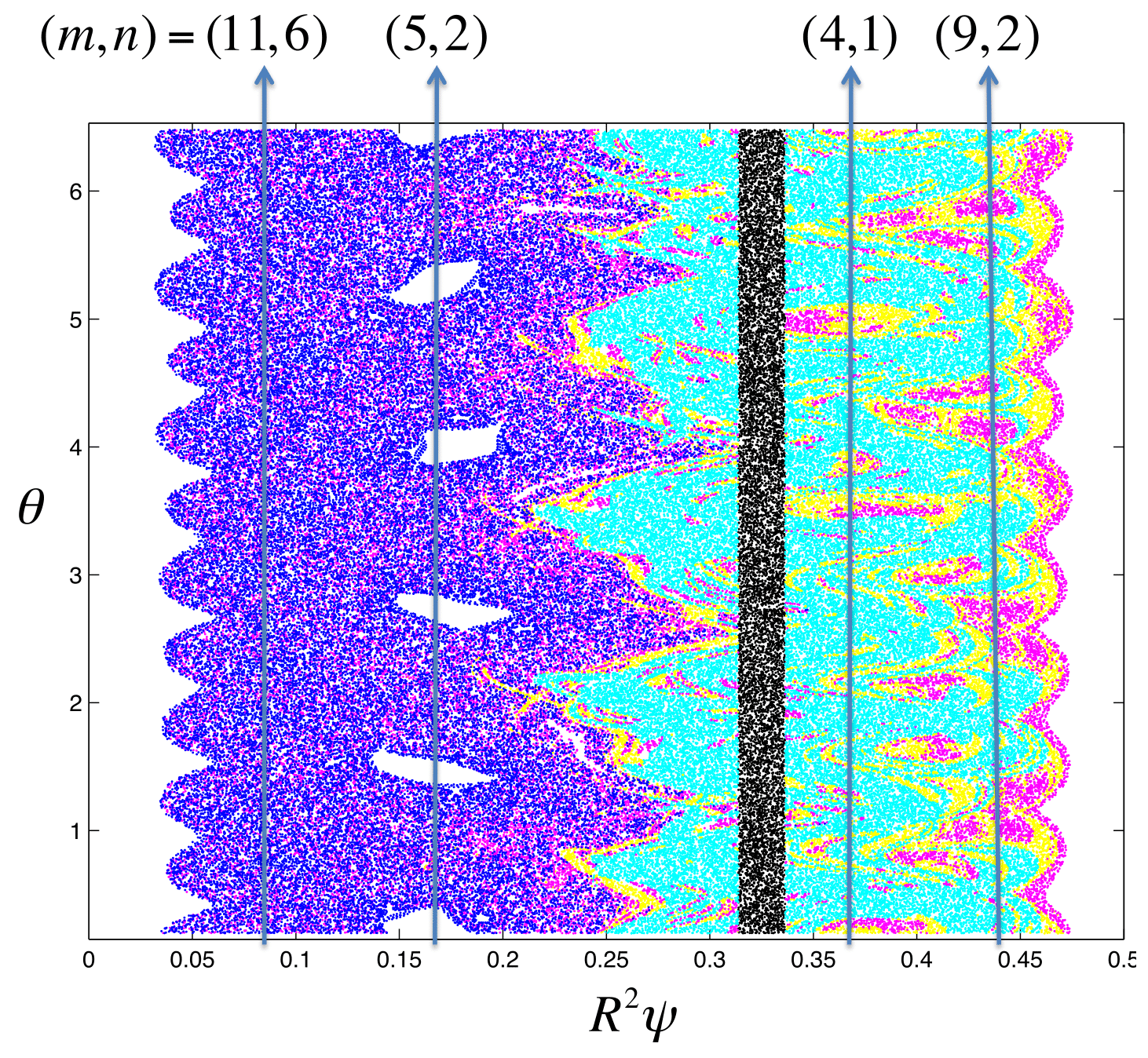}
\includegraphics[width=0.50 \columnwidth]{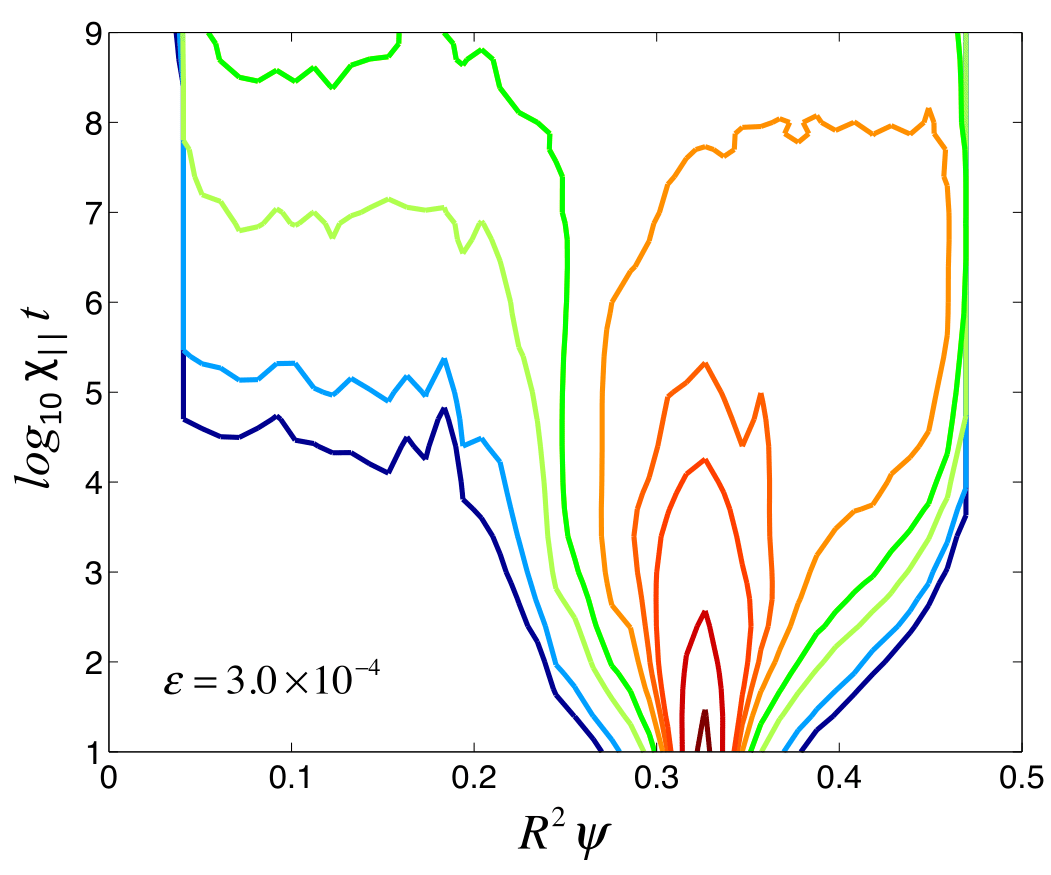}
\caption{Same as Fig.~\ref{fig_CL_count_1_5} but for $\epsilon=3 \times 10^{-4}$.}
\label{fig_CL_count_3_0}
\end{figure}

To study the role of weak magnetic field stochasticity on transport, we considered an equilibrium with monotonic $q$-profile in Eq.~(\ref{q_monotonic}),
perturbed by four overlapping modes with
\bq
\label{four_modes}
(m,n)=\left \{  (11, 6), (5, 2), (9, 2), (4, 1) \right\}\, ,
\eq
 and performed numerical simulations with 
increasing values of $\epsilon$, namely $\epsilon=1.5 \times 10^{-4}$, $\epsilon=2.0 \times 10^{-4}$ , $\epsilon=2.5 \times 10^{-4}$  and $\epsilon=3.0 \times 10^{-4}$. 
In all these  calculations, $B_z=1$, $B=0.09$, $R=5$, and $\lambda=0.57$,  and $q_0=1.23$.  
The set of modes in Eq.~(\ref{four_modes}) was selected so that the magnetic field exhibits weak stochasticity for the chosen values of $\epsilon$. 
This perturbation is interesting due to the absence of magnetic flux surfaces, and the presence of high order resonances in the middle of the domain which, as we will study below,  have a nontrivial 
effect on heat transport. 
Figures~\ref{fig_CL_count_1_5}-\ref{fig_CL_count_3_0} show the corresponding Poincare plots, indicating the spatial location of the resonances, obtained from the numerical integration of a single  magnetic field initial condition for a large number of crossings. 
Note that in the Poincare plots, as well as in  all the subsequent plots, we use $\psi$ defined  in Eq.~(\ref{psi_def}), as the radial coordinate.
It is observed that in all cases there are no flux-surfaces, i.e. there are 
no barriers to radial transport in the interval $\psi \in (0.075, 0.45)$. 
However, there is a significant number of  stability islands. 
We refer to this regime as ``weakly chaotic" to distinguish it from the fully chaotic regime in which there are no stability islands. As expected,  
increasing $\epsilon$ leads to the growth of the chaotic region and the shrinking of the stability islands.  The Poincare plots Figs.~\ref{fig_CL_count_1_5}-\ref{fig_CL_count_3_0} are color coded by the magnetic field connection length, $\ell_{B}$, which will be discussed below. 

\begin{figure}
\includegraphics[width=0.45 \columnwidth]{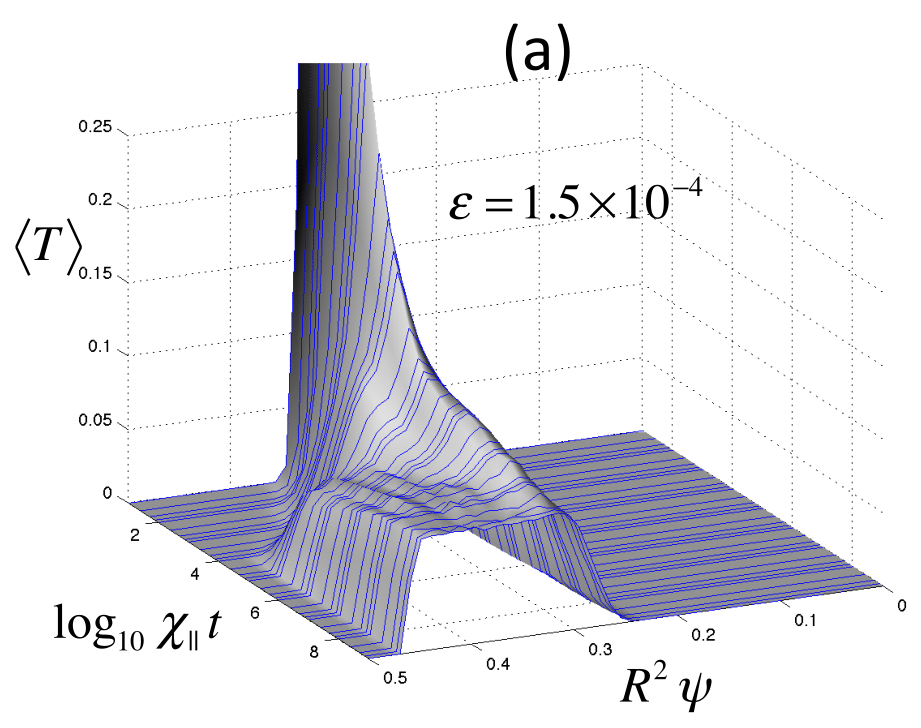}
\includegraphics[width=0.45 \columnwidth]{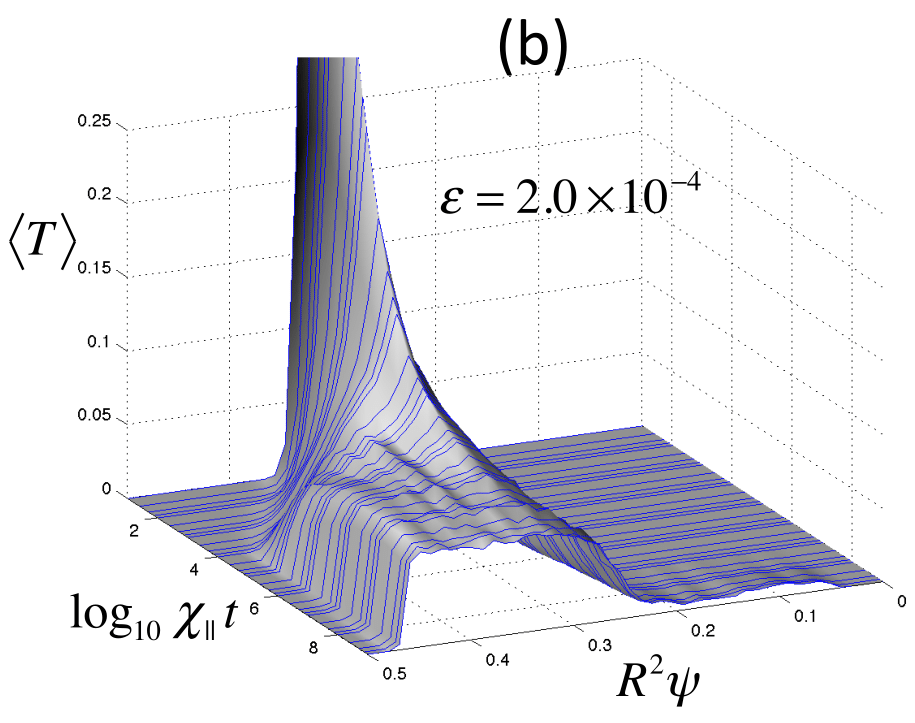}
\includegraphics[width=0.45 \columnwidth]{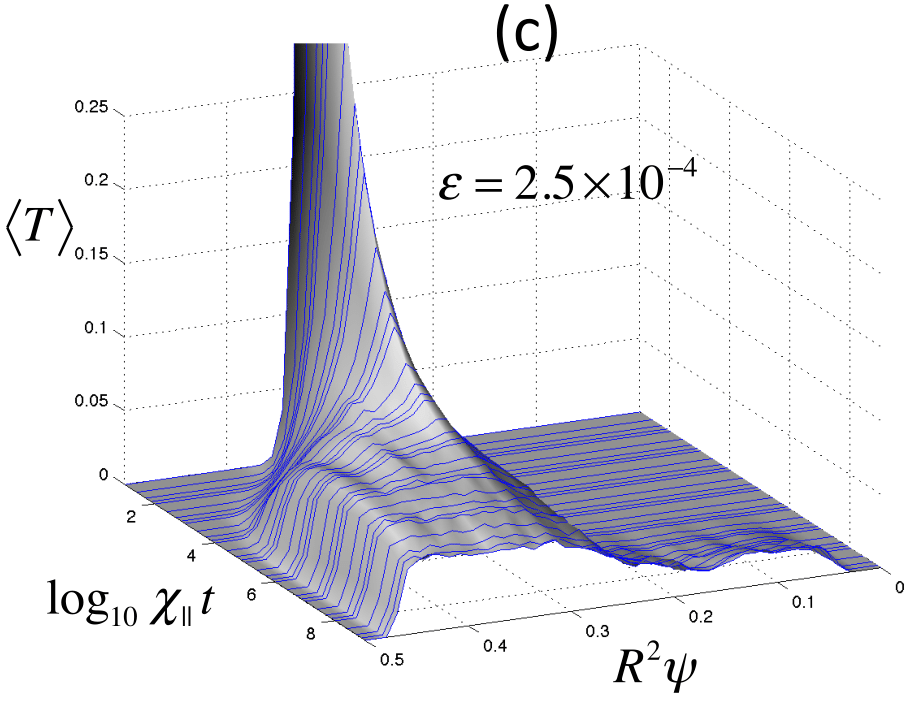}
\includegraphics[width=0.45 \columnwidth]{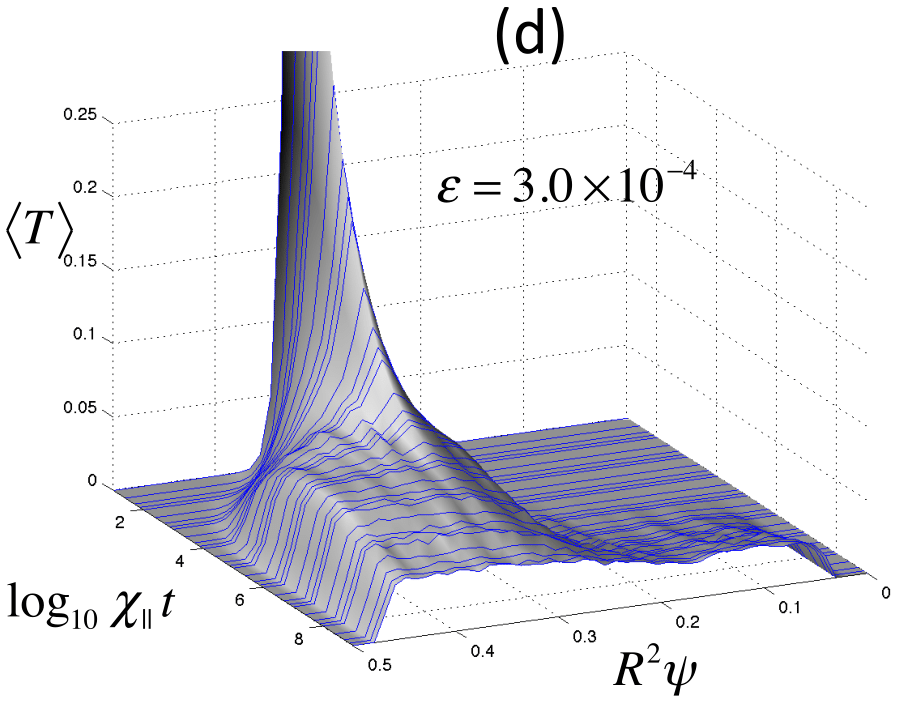}
\caption{(Color online)
Dependence of heat pulse propagation on magnetic field stochasticity. The 
magnetic field has the monotonic $q$-profile 
in Eqs.~(\ref{q_monotonic}), and the perturbation in Eq.~(\ref{eq_21}) includes the four modes in Eq.~(\ref{four_modes}).
The four panels show the result for different values of  perturbation amplitude $\epsilon$.
The initial condition corresponds to the temperature distribution in Eq.~(\ref{ic_T}). }
\label{fig_T_3D_a}
\end{figure}

Figure~\ref{fig_T_3D_a} shows the spatio-temporal evolution of  
$\langle T \rangle$.
As the Poincare sections in Figs.~\ref{fig_CL_count_1_5}-\ref{fig_CL_count_3_0} show, the level of stochasticity of the magnetic field varies with $\epsilon$, and this reflects on the radial propagation and penetration of the heat pulse. In all cases the propagation of the pulse in the edge region, $R^2 \psi \in (0.325, 0.5)$, exhibits a similar behavior. This is consistent with the Poincare plots that show a similar level of stochasticity in that region for the four values of $\epsilon$ considered. However, the inwards propagation of the pulse shows a strong dependence on $\epsilon$. For the smallest value,  $\epsilon = 1.5 \times 10^{-4}$, Fig.~\ref{fig_T_3D_a}-(a) shows that the pulse practically does not penetrate beyond $R^2 \psi \sim 0.275$. This is again consistent with the Poincare plot in Fig.~\ref{fig_CL_count_1_5} that shows a significant number of stability islands located near $R^2 \psi \sim 0.25$. 
On the hand, the reduction of the integrability regions with increasing 
$\epsilon$ shown in the Poincare plots explains the increase of the penetration depth of the heat pulse 
as well as the increase of $\langle T \rangle$ near the core for  large values of $\epsilon$.

To gain further understanding on the role of magnetic field stochasticity on heat transport,
we compare the connection length of the magnetic field with the radial penetration of the heat pulse. 
Given a region, ${\cal R}$, in the $(\psi,\theta)$ plane and a point, $P_i$, in the Poincare section, the connection length,  $\ell_{B}$, is determined by the number of iterations to go from $P_i$ to ${\cal R}$ and from ${\cal R}$ to $P_i$.  
More precisely, 
given a point $\left( \psi_i, \theta_i \right)$ in the Poincare section 
$\left \{ \left( \psi_0, \theta_0 \right),  \left( \psi_1, \theta_1 \right), \ldots  \left( \psi_N, \theta_N \right) \right \}$ (where $\left( \psi_k, \theta_k\right)$ for $k=1,\, \ldots N$ denotes the $k$-th crossing starting from the initial condition $\left( \psi_0, \theta_0 \right)$) we search for the smallest integer $n_+$ such that 
$\left( \psi_{i+n_+}, \theta_{i+n_+} \right) \in {\cal R}$ and the smallest integer $n_-$ such that 
$\left( \psi_{i-n_-}, \theta_{i-n_-} \right) \in {\cal R}$ and define
\bq
\label{cl}
\ell_B = \min \{ n_+, n_-\} \, . 
\eq
If $n_+$ exists but $n_-$ does not exist, we define $\ell_B =n_+$.
Conversely, if  $n_-$ exists but $n_+$ does not exist, $\ell_B =n_-$. In the event when neither $n_+$ nor $n_-$ exist, $\ell_B= \infty$.  Since we are interested in the propagation of radially localized pulses we define ${\cal R}$ as 
\bq
\label{region}
{\cal R} = \left \{ \left( \psi, \theta \right) \left | \right. \psi_1 < \psi < \psi_2 \right \}  \, ,
\eq
where $\psi \in (\psi_1, \psi_2)$ is the region where the heat pulse is initially concentrated. 
According to Eq.~(\ref{ic_T}), the initial condition is a Gaussian distribution 
centered at $R \psi_0=0.25$ with $\sigma_p =0.008$. Based on this, we set
$R \psi_1=0.314$ and $R \psi_2=0.336$ which defines an interval of width $R \psi_2-R \psi_1 \sim 3 \sigma$
centered at $R \psi_0=0.25$. 

\begin{figure}
\includegraphics[width=0.75 \columnwidth]{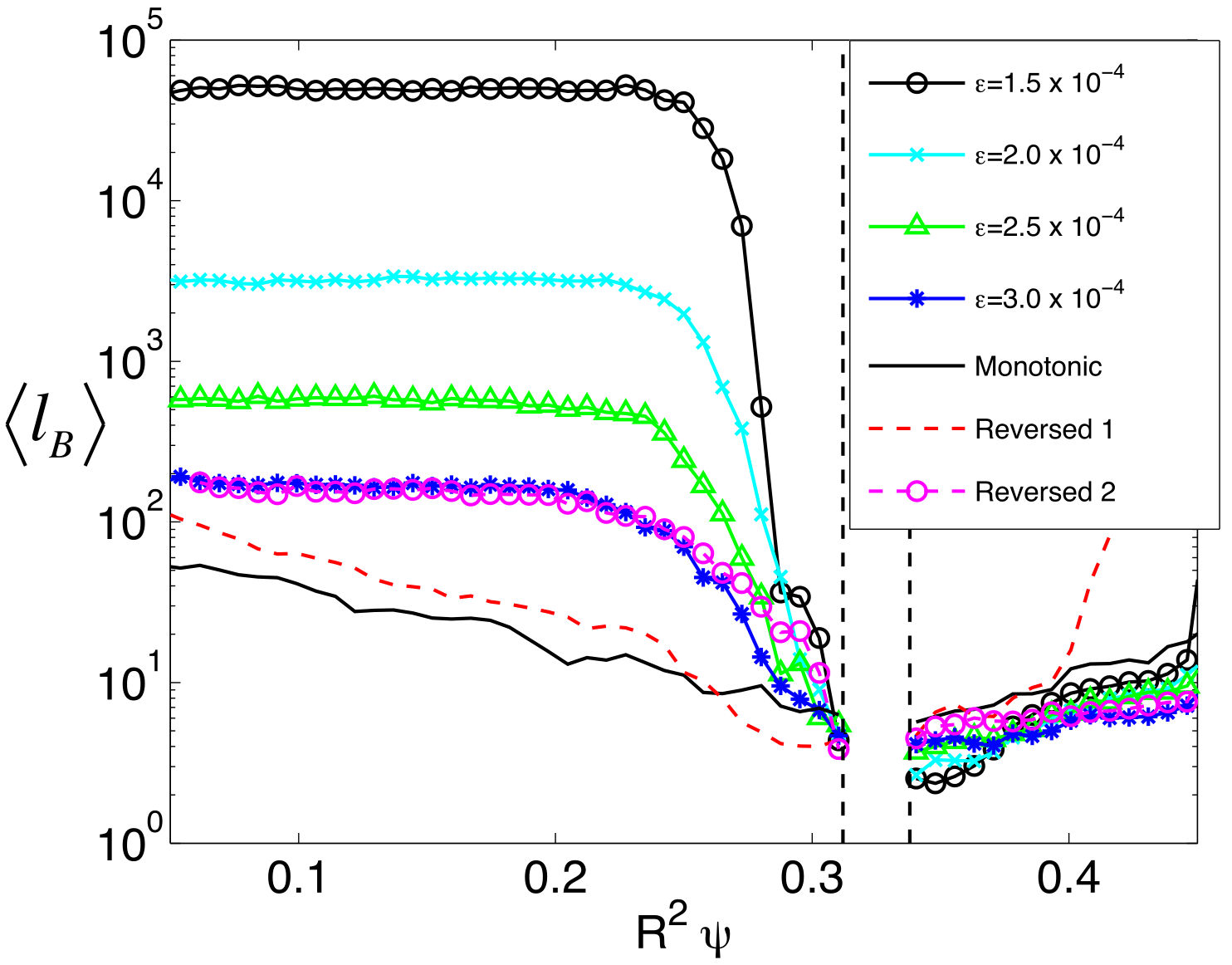}
\caption{(Color online) Magnetic field connection length averaged on $\theta$, $\langle \ell_B \rangle$, as function of 
$\psi$ for different levels of magnetic field stochasticity, and different $q$-profile configurations. 
The results labeled as 
$\epsilon=1.5 \times 10^{-4},\, 2.0 \times 10^{-4},\, 2.5 \times 10^{-4}$ and $3.0 \times 10^{-4}$
correspond to the weakly chaotic, monotonic $q$ runs in Figs.~\ref{fig_CL_count_1_5},
\ref{fig_CL_count_2_0}, \ref{fig_CL_count_2_5} and ~\ref{fig_CL_count_3_0} respectively. 
The result labeled ``Monotonic" corresponds to the fully chaotic, monotonic $q$ case in Fig.~\ref{fig_CL_count_twist}. 
The two reversed shear cases, ``Reversed 1" and ``Reversed 2", correspond to the fully chaotic, reversed shear cases  in Figs.~\ref{fig_CL_count_nontwist} and ~\ref{fig_CL_count_nontwist_2} respectively. The interval $\psi \in \left(0.31, 0.34\right)$ denotes the region where the heat pulse is introduced.
} 
\label{fig_CL_vs_psi}
\end{figure}

The top panels of Figs~\ref{fig_CL_count_1_5}-\ref{fig_CL_count_3_0} show the connection lengths of the weakly chaotic magnetic field 
with monotonic $q$--profile. For each value of $\epsilon$, the plots were generated by color-coding (using the value of 
$\ell_B$) each point of the corresponding Poincare section generated from a single initial condition. The black rectangular strip corresponds to the region ${\cal R}$ and therefore, by definition, $\ell_B=0$ there. 
It is observed that the distribution and values of the connection length in the ``edge" region, $R^2 \psi \in (0.0336, 0.45)$, is more or less similar for all the values of $\epsilon$. However a very strong dependence of $\ell_B$ on $\epsilon$ is observed elsewhere
in the ``core"   region, $R^2 \psi \in (0.05, 0.314)$. 
This is consistent with the existence of stability islands generated by the higher order resonances of the chaotic fields for small values of 
$\epsilon$. The high-order resonances  around $R^2 \psi \sim 0.275$, 
along with the potential existence of Cantori
\cite{cantori,hudson}  there, play a critical role.  
In fact, as shown in Fig.~\ref{fig_CL_vs_psi},  in the
``intermediate" region $R^2 \psi \in (0.225, 0.314)$ the 
averaged in $\theta$ connection length, $\langle \ell_B \rangle$, exhibits its sharpest gradient. This figure also indicates that $\langle \ell_B \rangle$ is approximately   constant in the 
``core" region $R^2 \psi \in (0.05, 0.225)$. The key issue is the close to exponential growth of the average connection length
in the core region, $\langle \ell_B \rangle=\{285, 594, 3387, 52513\}$, resulting from the moderate decrease of the 
magnetic field perturbation, $\epsilon=\{ 3.0 \times 10^{-4}, 2.5 \times 10^{-4}, 2.0 \times 10^{-4}, 1.5\times 10^{-4} \}$. 

There is a close relationship between the  magnetic field connection length,  $\langle \ell_B \rangle$, and the transport properties of 
localized heat pulses in the LG method. 
In particular, as discussed in Sec.~II, according to Eq.~(\ref{eq_II_10}) the value of $T$ at a time $t$ at ${\bf r}_0$ is obtained by summing the contributions of the initial condition  $T_0$ along the corresponding magnetic field line path 
${\bf r}={\bf r}(s)$ with ${\bf r}(s=0)={\bf r}_0$. In the case of a localized pulse initial condition, the magnetic field line going though
 ${\bf r}_0$ will in general pierce the region ${\cal R}$ in Eq.~(\ref{region}) one or several times, and each interesection will add a contribution to the temperature
 at ${\bf r}_0$. Due to the rapid decay of the diffusive Green's function in Eq.~(\ref{eq_II_12}) the largest contribution to the temperature will come from the closest intersection with ${\cal R}$, and the strength of the contribution will depend on the distance along the field line between 
 ${\bf r}_0$ and ${\cal R}$, which is directly related to the magnetic field connection length as defined in Eq.~(\ref{cl}). From this it follows that the temperature response at a point ${\bf r}_0$ is a monotonically increasing function of the magnetic field connection length at ${\bf r}_0$. 

The  relationship between $\langle \ell_B \rangle$ and the temperature response 
provides valuable insights in the 
understanding of  the  
dependence of the radial penetration depth of heat pulses on the level of stochasticity of the magnetic field. 
In particular, as Fig.~\ref{fig_CL_count_1_5} shows, for $\epsilon=1.5 \times 10^{-4}$  the level of stochasticity is relatively small and as a result the connection length in the core region is quite large, $\ell_{B} \sim 5 \times 10^4$. This explains why, as 
shown in the bottom panel of  Fig.~\ref{fig_CL_count_1_5},
the heat pulse evolves as if there is transport barrier around $R^2 \psi \sim 0.25$ at it does not penetrate to the core. 
The contour plots shown for successively increasing values of $\epsilon$ in Figs.~\ref{fig_CL_count_2_0}-\ref{fig_CL_count_3_0},
show how an increase of $\epsilon$ leads to a decrease of $\ell_B$ and a larger radial penetration of the heat pulse. 

\begin{figure}
\includegraphics[width=0.75 \columnwidth]{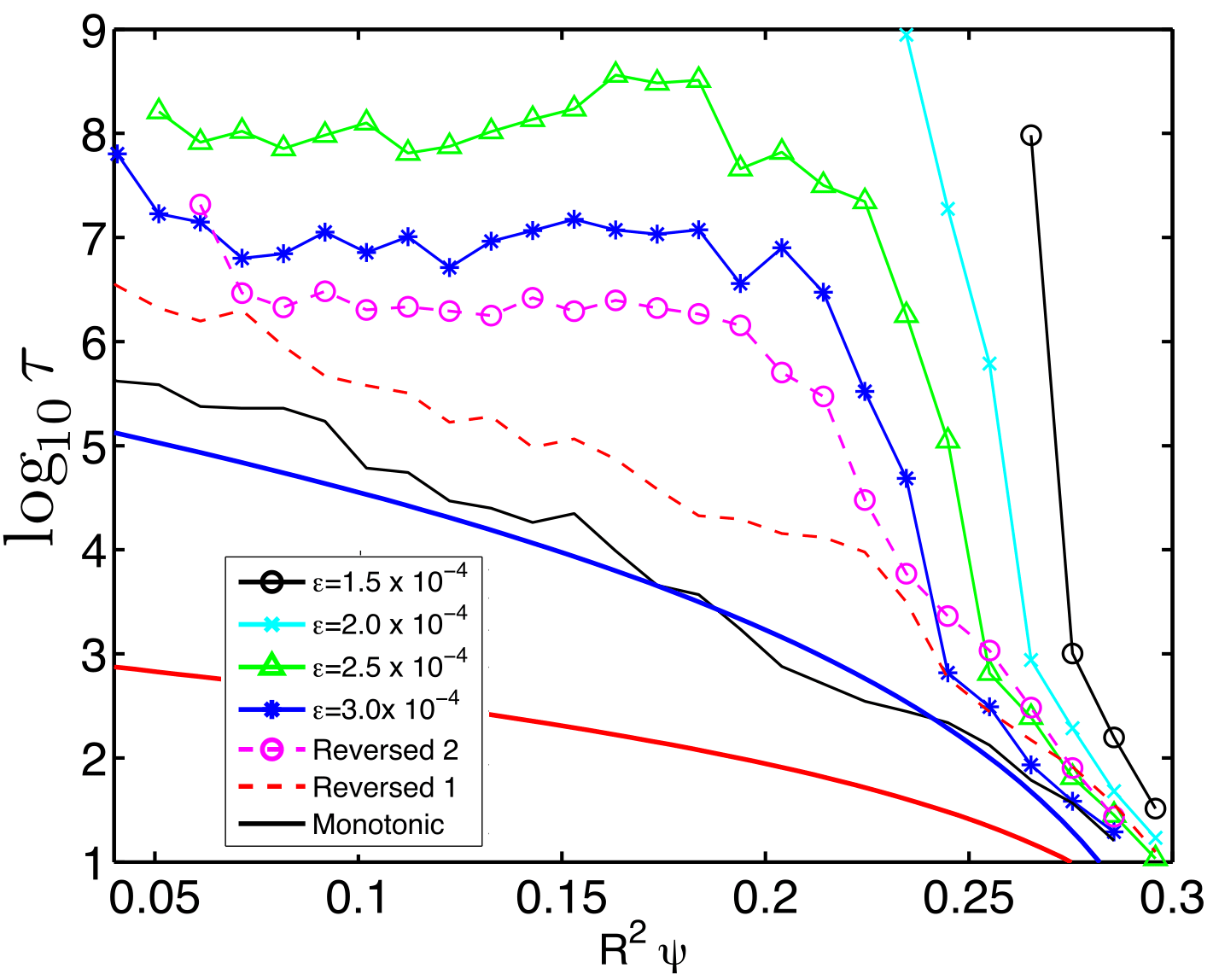}
\caption{(Color online) Time delay of temperature response at different $\psi$ locations.
The results labeled as 
$\epsilon=1.5 \times 10^{-4},\, 2.0 \times 10^{-4},\, 2.5 \times 10^{-4}$ and $3.0 \times 10^{-4}$
correspond to the weakly chaotic, monotonic $q$ runs in Figs.~\ref{fig_CL_count_1_5},
\ref{fig_CL_count_2_0}, \ref{fig_CL_count_2_5} and ~\ref{fig_CL_count_3_0} respectively. 
The result labeled ``Monotonic" corresponds to the fully chaotic, monotonic $q$ case in Fig.~\ref{fig_CL_count_twist}. 
The two reversed shear cases, ``Reversed 1" and ``Reversed 2", correspond to the fully chaotic, reversed shear cases  in Figs.~\ref{fig_CL_count_nontwist} and ~\ref{fig_CL_count_nontwist_2} respectively.
The two, smooth solid line curves in the lower part of the plot correspond, from top to bottom, to the
sub-diffusive and diffusive delay according to the self-similar model in Eq.~(\ref{self_sim_T}).
 }
\label{fig_delays}
\end{figure}

A commonly used diagnostic in the study of perturbative transport are the temperature traces as functions of time at different radial locations. Of particular interest is the delay time, $\tau$, of the temperature response. For a given temperature threshold, $T_*$, the delay time $\tau$ at a radial location $\psi$ is defined by the condition $\langle T \rangle (\psi, \tau) = T_*$. Figure~\ref{fig_delays} shows $\tau$ for a threshold $T_*=10^{-2}$ as function of $\psi$ for all the magnetic field configurations considered. In the weakly chaotic cases, $\tau$ decreases when $\epsilon$ increases, and for  $\epsilon=1.5 \times 10^{-4}$ and $\epsilon=2.0 \times 10^{-4}$ there is no response for $R^2 \psi < 0.225$ consistent with the fact that as shown in Figs.~\ref{fig_CL_count_1_5}-\ref{fig_CL_count_2_0} in these cases the $T=T_*$ contours do not penetrate the core. The strongest gradient in the delay time occurs around $R^2 \psi \sim 0.25$, the same region where the connection length changes most rapidly and where the radial penetration of the pulse slows down. 

\begin{figure}
\includegraphics[width=0.6 \columnwidth]{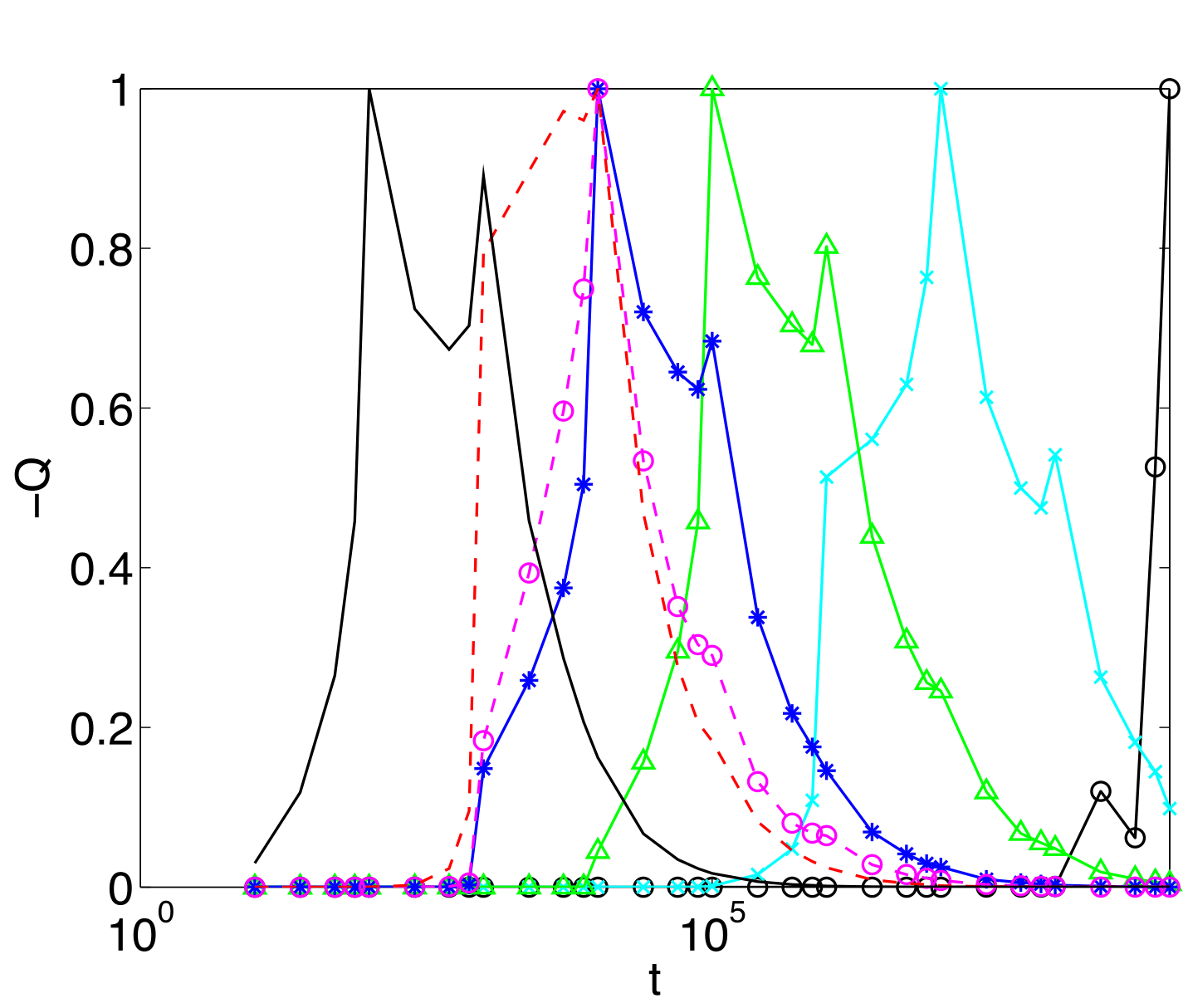}
\includegraphics[width=0.6 \columnwidth]{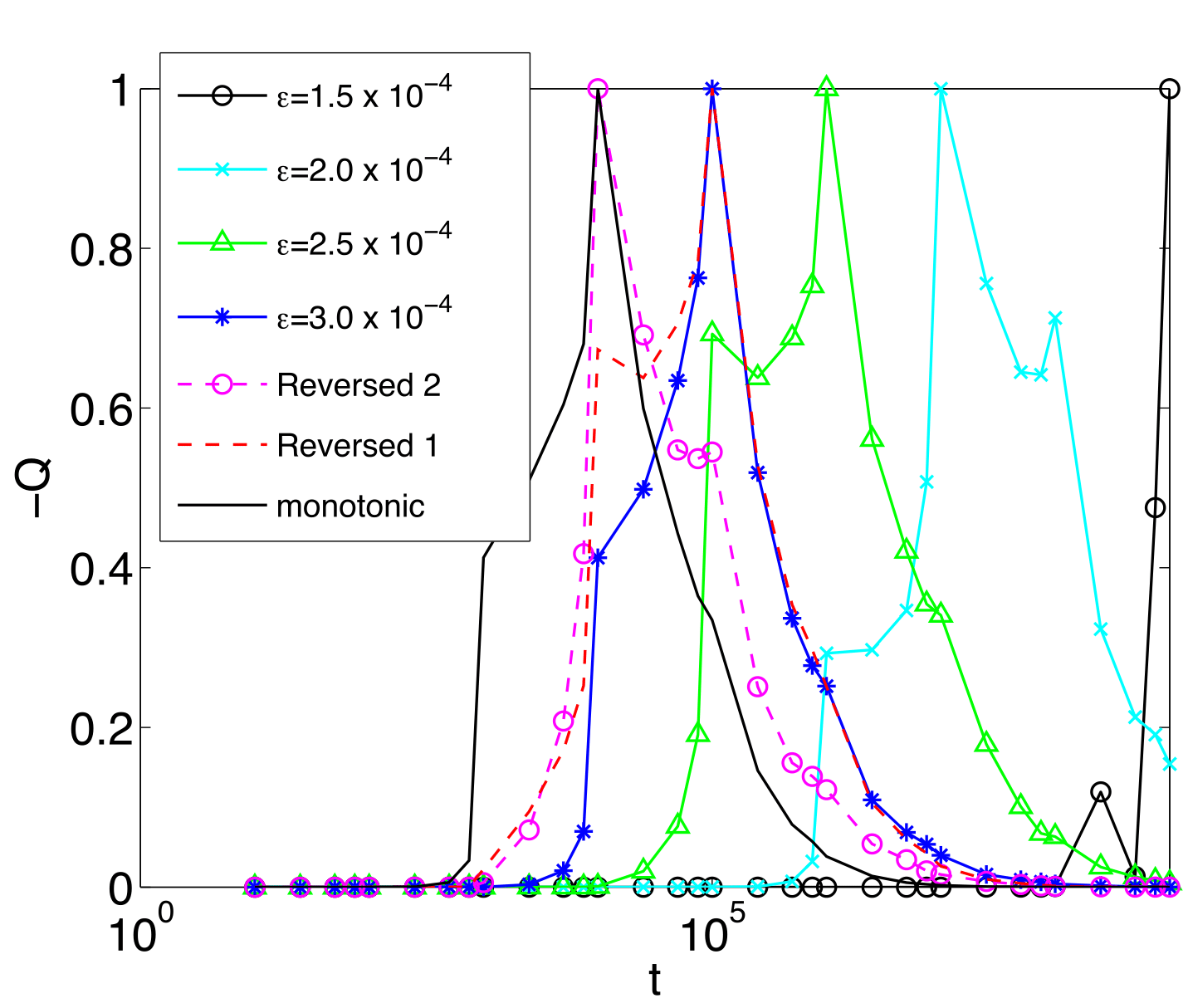}
\caption{(Color online) Normalized heat flux in Eq.~(\ref{flux_psi}) at a fixed $\psi$, as function of time. The top panel corresponds to $\psi=0.2$ and the bottom panel to $\psi=0.1$. For visualization purposes, in each case the negative of the flux, $-Q$, has been plotted, after  been normalized by its maximum absolute value, 
$q_{max}$. 
The results labeled as 
$\epsilon=1.5 \times 10^{-4},\, 2.0 \times 10^{-4},\, 2.5 \times 10^{-4}$ and $3.0 \times 10^{-4}$
correspond to the weakly chaotic, monotonic $q$ runs in Figs.~\ref{fig_CL_count_1_5},
\ref{fig_CL_count_2_0}, \ref{fig_CL_count_2_5} and ~\ref{fig_CL_count_3_0} respectively. 
The result labeled ``Monotonic" corresponds to the fully chaotic, monotonic $q$ case in Fig.~\ref{fig_CL_count_twist}. 
The two reversed shear cases, ``Reversed 1" and ``Reversed 2", correspond to the fully chaotic, reversed shear cases  in Figs.~\ref{fig_CL_count_nontwist} and ~\ref{fig_CL_count_nontwist_2} respectively.
 }
\label{fig_fluxes}
\end{figure}

The dependence of the radial heat flux, 
$\langle{{\bf q}\cdot {\hat e}_\psi} \rangle$,
on the level of magnetic field stochasticity is also interesting.  
From the continuity equation,  it follows that
\bq 
\label{flux_psi}
\langle
{{\bf q}\cdot {\hat e}_\psi} \rangle=-\frac{1}{\sqrt{2 \psi}}\,
\frac{d}{dt} \int_0^\psi \langle T \rangle d \psi' \, .   
\eq 
Figure~\ref{fig_fluxes} shows the fluxes as function of $t$ for  $R^2 \psi=0.1$ and $R^2 \psi=0.2$ computed directlyy from Eq.~(\ref{flux_psi}). To ease the comparison, the flux values have been normalized in each case by the  maximum of the absolute value of the flux. In all cases, the flux exhibits a single minimum at a time that increases with the decrease of the level of stochasticity. The values of the absolute value of the flux at $R^2 \psi=0.2$ show a strong dependence on $\epsilon$ in the weakly chaotic cases. 
In particular the corresponding values of $\epsilon$, and  $q_{max}$ there, are:
$( \epsilon,  q_{max})=
(1.5 \times 10^{-4},  1.3 \times 10^{-16}),
(2.0 \times 10^{-4}, 4.0 \times 10^{-12}), 
(2.5 \times 10^{-4}, 1.4 \times 10^{-10})$, and 
$(3 \times 10^{-4},2.4 \times 10^{-9} )$.

\subsection{Strongly chaotic field in monotonic $q$-profile}
\label{strong_twist}

\begin{figure}
\includegraphics[width=0.55\columnwidth]{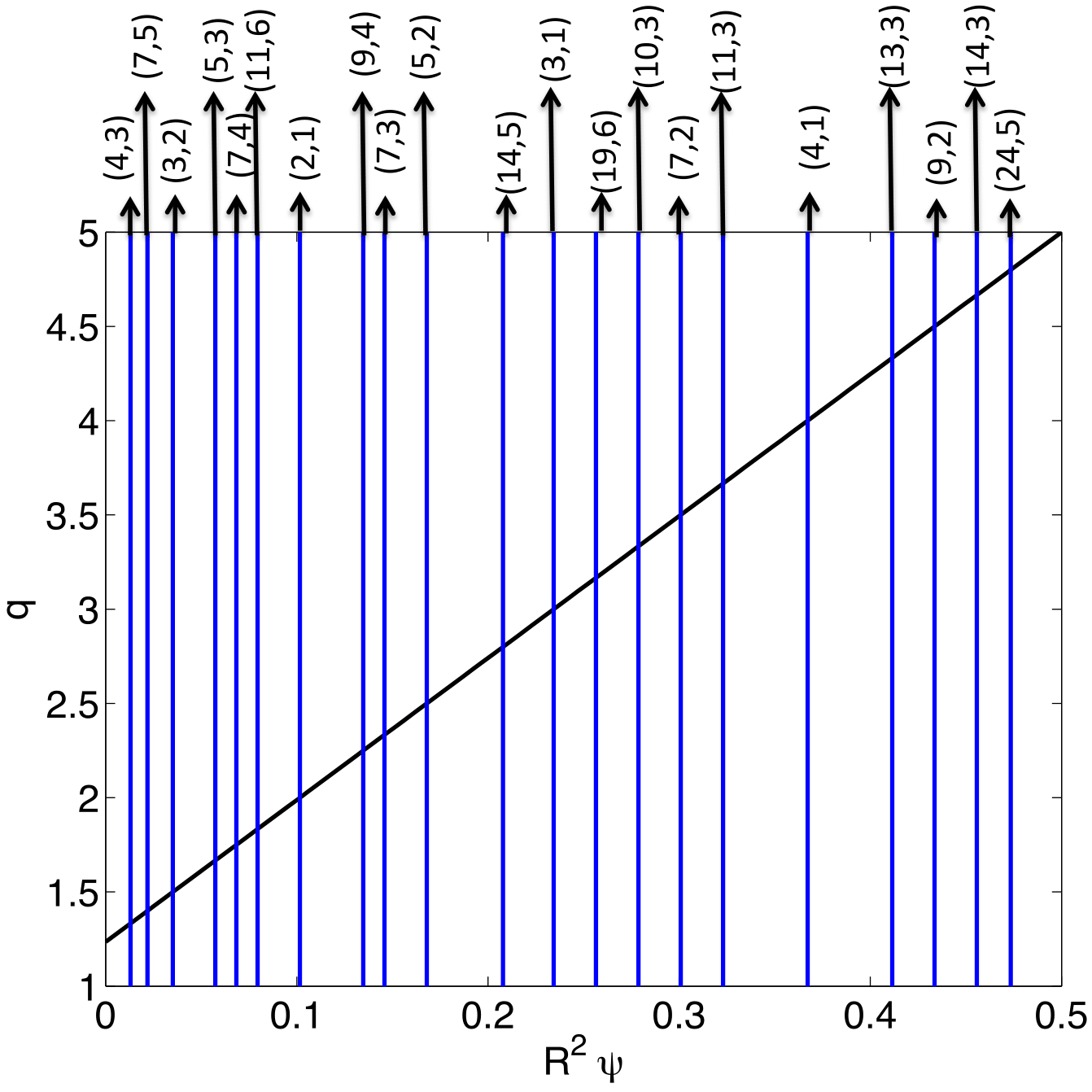}
\caption{(Color online) $q$-profile and resonant modes used in the numerical study of heat transport in fully chaotic magnetic fields in a monotonic $q$ 
configuration. 
The solid line shows the $q$-profile in Eq.~(\ref{q_monotonic}) as function of $R^2 \psi$.
The modes are given in Eq.~(\ref{modes}), and the vertical lines indicate the corresponding radial location of the resonances.
}
\label{qprof_monotonic}
\end{figure}

To study the propagation of heat pulses in strongly chaotic fields, we consider the monotonic $q$-profile 
equilibrium in Eq.~(\ref{q_monotonic}) perturbed by 
 twenty overlapping modes with 
 \begin{equation} 
\label{modes}\begin{split}
& (m,n) = \{ (4,3), (7,5), (3,2), (5, 3), (7,4),\\
& (11,6), (2,1), (9,4), (7, 3), (5,2), (14,5),\\
&  (3,1), (19,6), (10,3), (7,2),(11,3), (4,1),\\
&   (13,3), (9,2), (14,3), (24,5)\} \, .
 \end{split}\end{equation}
The  location of the resonances created by these modes is shown in Fig.~\ref{qprof_monotonic}. As observed in the figure, the chosen values of $(m,n)$ 
give rise to an approximately uniform distribution of resonances in the  radial-$\psi$ domain. 
To decorrelate the modes, the phases of the perturbations, 
$\{ \zeta_{mn}\}$, in Eq.~(\ref{perturbation})  were chosen from a random distribution on the interval $[0, 2 \pi]$. In this case the perturbation amplitude is $\epsilon=10^{-4}$, and the width of the perturbation
function in Eq.~(\ref{eq_21})  is $\sigma=0.5$. 

\begin{figure}
\includegraphics[width=0.53 \columnwidth]{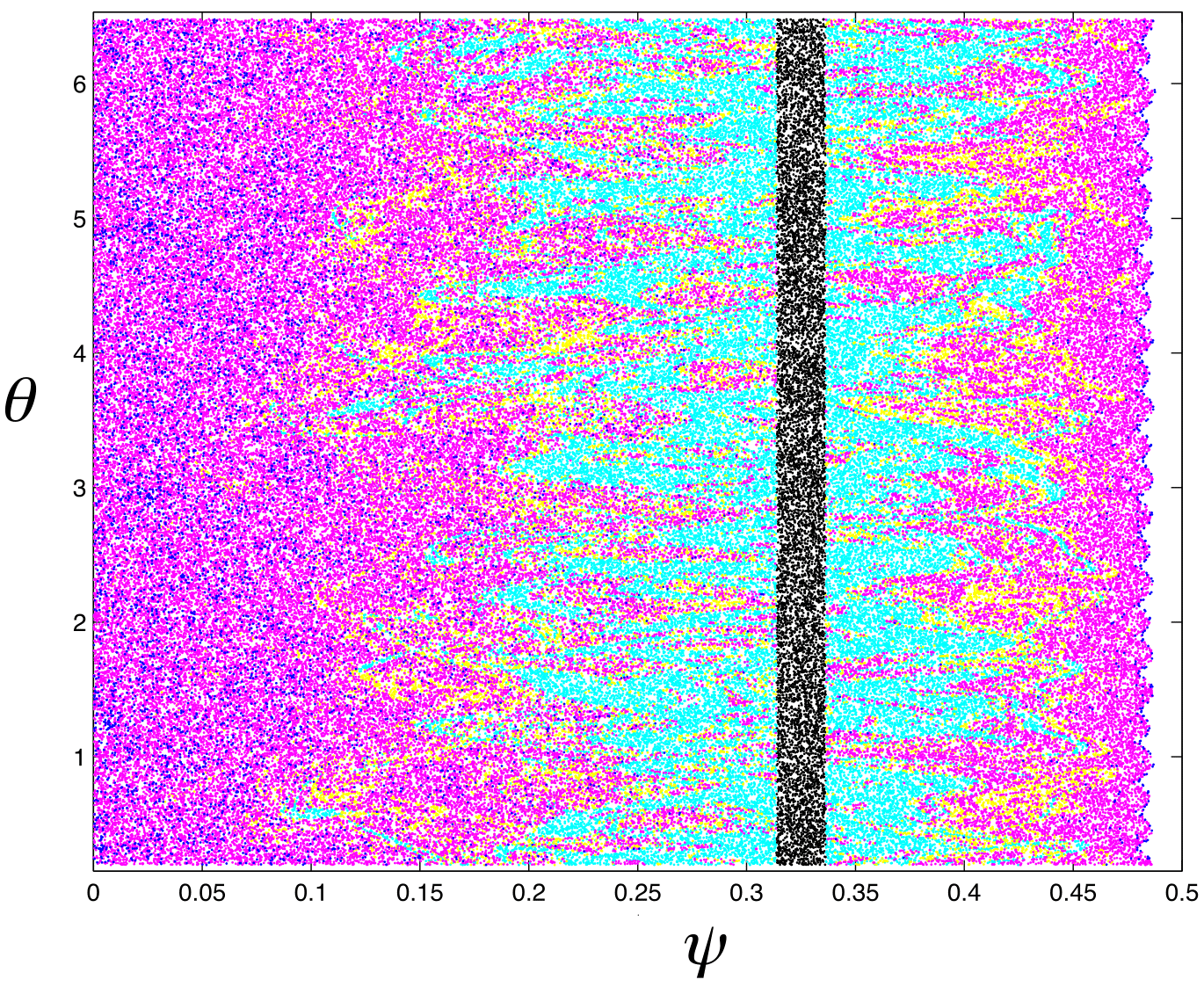}
\includegraphics[width=0.55 \columnwidth]{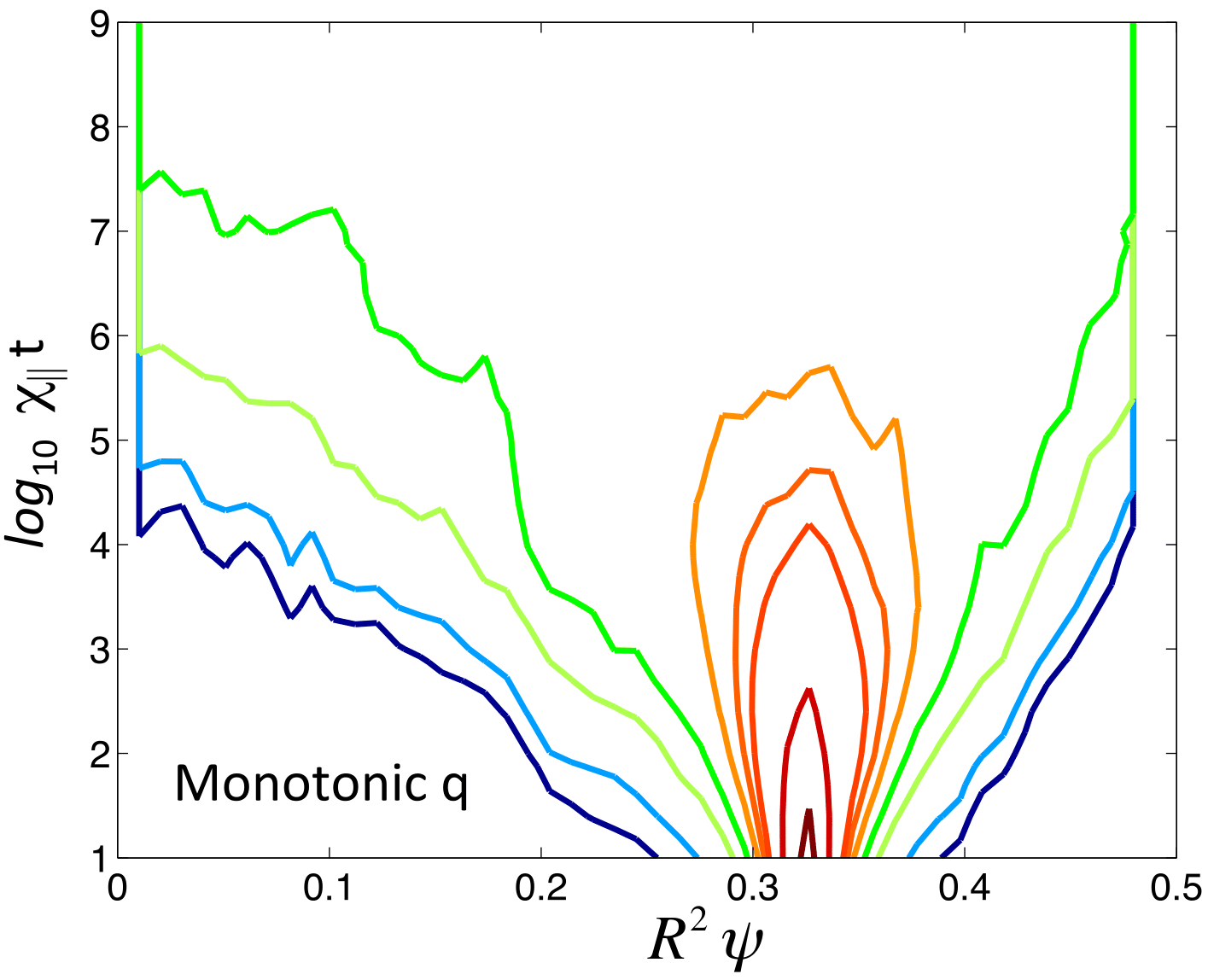}
\caption{
Magnetic field connection length and heat pulse propagation 
in a {\em strongly chaotic} (no flux surfaces) magnetic field with {\em monotonic $q$ profile}. 
Top panel shows a Poincare plot  of the magnetic field with $q$-profile in 
Eq.~(\ref{q_monotonic})  perturbed by the twenty one overlapping modes in Eq.~(\ref{modes}) 
with $\epsilon=1\times 10^{-4}$ and $\sigma=0.5$. The points 
correspond to the iterates of a single initial condition, 
 color coded by the value of magnetic field connection length, $\ell_B$, defined in Eq.(\ref{cl}). 
The color scale is the same as that used in Figs.~\ref{fig_CL_count_1_5}-\ref{fig_CL_count_3_0}.
The corresponding $q$-profile with the location of the resonances is shown in Fig.~\ref{qprof_monotonic}.
The contour plot in the bottom panel shows the spatio-temporal evolution of the heat pulse. Like in Figs.~\ref{fig_CL_count_1_5}-\ref{fig_CL_count_3_0}, the contours correspond to 
(from red to blue) $T_0=\{ 0.5,\, 0.25,\, 0.10,\, 0.075,\,  0.05,\, 0.025,\, 10^{-2},\, 10^{-3},\,  10^{-4} \}$.}
\label{fig_CL_count_twist}
\end{figure}

Figure~\ref{fig_CL_count_twist} shows the Poincare plots obtained from the numerical integration of a single  magnetic field initial condition for a large number of crossings. As expected, no flux surfaces or magnetic islands are observed in the fully chaotic Poincare section. 
Like in the previously discussed weakly chaotic cases, the
plot is color coded by the magnetic field connection length, $\ell_{B}$. 
As shown in Fig.~\ref{fig_CL_vs_psi}, the 
averaged connection length is significantly smaller than the 
averaged connection length in the weakly chaotic cases, and this reflects in the significantly faster and deeper radial penetration of the heat pulse as observed in the bottom panel of Fig.~\ref{fig_CL_count_twist}.

In Refs.~\cite{DL,DL_pop} it was shown that, in the case of fully stochastic magnetic fields with monotonic $q$-profiles, the spatio-temporal evolution of a
localized temperature pulse follows the self-similar scaling
\bq 
\label{self_sim_T}
\langle T \rangle
(\psi,t) = \left( \chi_\parallel t \right)^{-\gamma/2} L_\alpha(\eta) \, , 
\eq 
where the similarity variable is defined as 
\bq
\eta=(\psi -\overline{\psi})/(\chi_\parallel t)^{\gamma/2} \, ,
\eq
$\gamma$ is the scaling exponent, and $L_\alpha$ is the scaling function. 
From  Eq.~(\ref{self_sim_T}) it follows that
the time evolution of the maximum of the temperature profile, $T_{max}$, and the temperature variance,
\bq
\label{Tm_Tsigma}
T_{max}=\langle T \rangle (0,t) \, , \qquad
\sigma_T=\overline  {\left( \psi - \overline{\psi} \right)}^2 (t) \, ,
\eq
where $\overline{f}=\int f \langle T \rangle d\psi$, scale as
$T_{max} \sim t^{-\gamma/2}$, and  $\sigma_T^2 \sim t^\gamma$. 

 The value of the scaling exponent $\gamma$, and the form of 
the scaling function, $L_\alpha$, depend on the physics of the parallel heat flux closure. 
In the case of parallel diffusive closure in Eq.~(\ref{eq_II_5}), which is the one of interest in the present paper, $\alpha=2$, 
 the scaling function
is an stretched exponential of the form 
\bq
\label{stretched_exp}
 L_2(\eta) =A \, e^{- |\eta/\mu |^\nu}  \, ,
\eq
and the scaling exponent $\gamma=1/2$ \cite{DL_pop}. This 
implies that the heat pulse exhibits the {\em sub-diffusive} 
scaling 
\bq
\label{moments_sca}
 T_{max} \sim t^{-1/4}\, , \qquad
 \sigma_T ^2 \sim t^{1/2} \, ,
 \eq
i.e., the spreading of the heat pulse is slower than the one expected from a diffusive transport process. 

\begin{figure}
\includegraphics[width=0.60 \columnwidth]{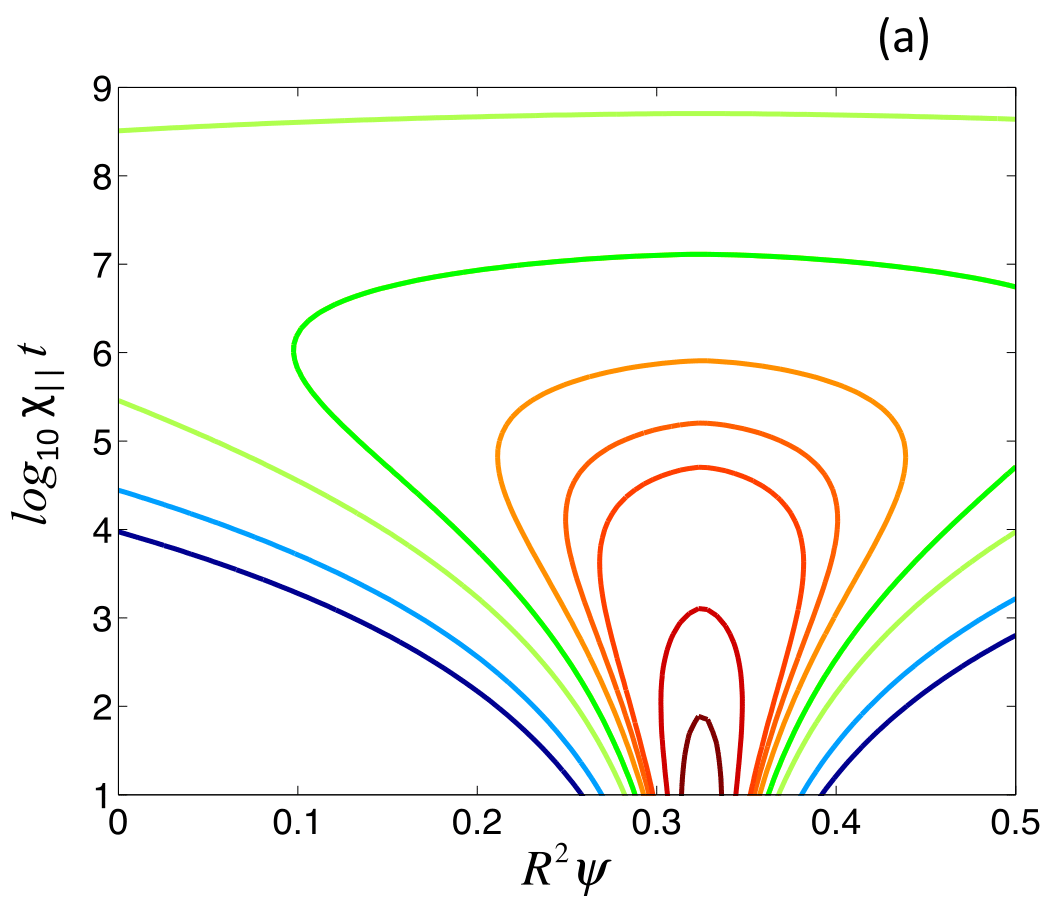}
\includegraphics[width=0.60 \columnwidth]{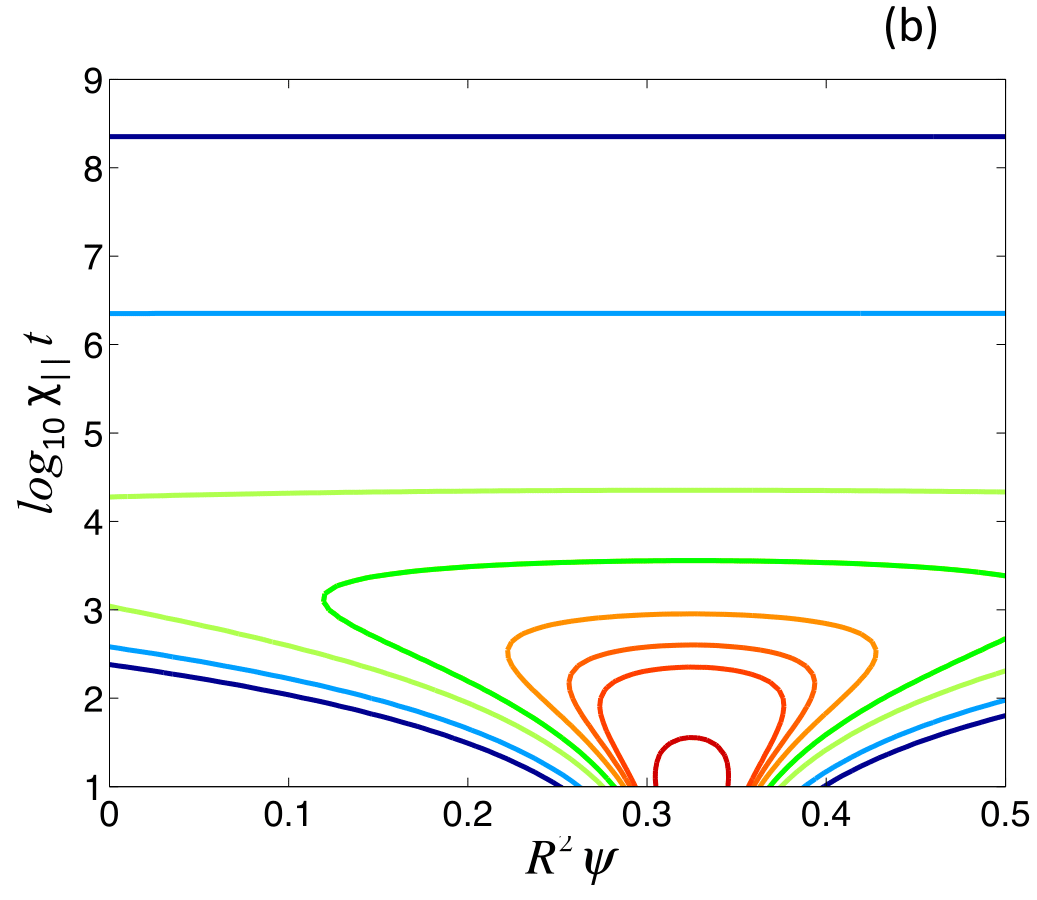}
\caption{(Color online) Spatio-temporal evolution of heat pulse according to the self-similar model in Eq.~(\ref{self_sim_T}). 
Panel (a) corresponds to the sub-diffusive scaling case with $\gamma=1/2$, $\nu=1.6$ and $\mu=0.0095$. 
Panel (b) corresponds to the diffusive scaling case with $\gamma=1$, 
$\nu=2$ and $\mu=0.008$. In both cases, the iso-contours correspond to the same values of those in Figs.~\ref{fig_CL_count_1_5}-\ref{fig_CL_count_nontwist}, namely 
(from red to blue) $T_0=\{ 0.5,\, 0.25,\, 0.10,\, 0.075,\,  0.05,\, 0.025,\, 10^{-2},\, 10^{-3},\,  10^{-4}$.}
\label{fig_count_self_sim}
\end{figure}

\begin{figure}
\includegraphics[width=0.75 \columnwidth]{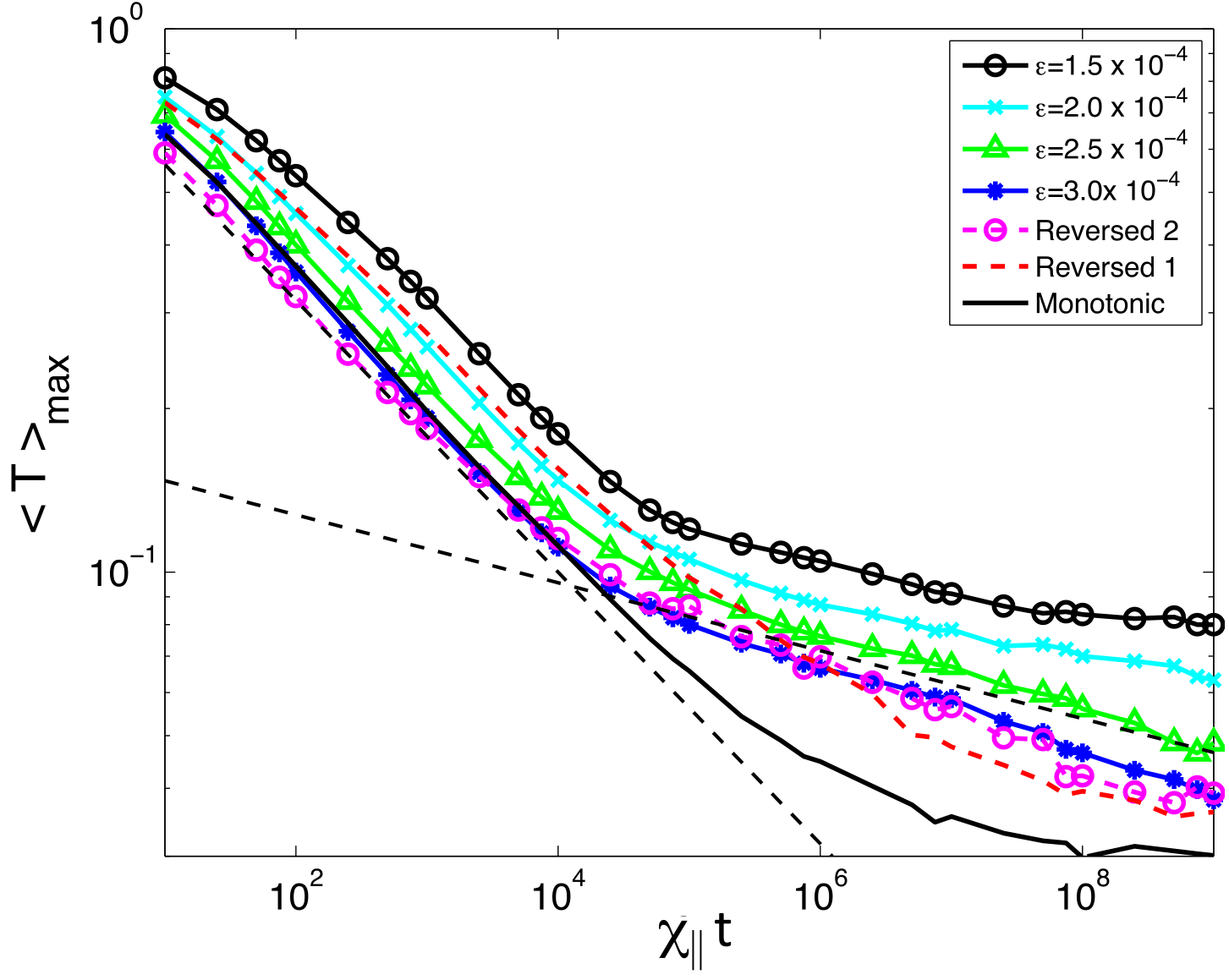}
\caption{(Color online)
Decay of maximum temperature,  
$\langle T \rangle_{max}$, as function of 
$\chi_\parallel t$ for different levels of magnetic field stochasticity, and different $q$-profile configurations. 
The results labeled as 
$\epsilon=1.5 \times 10^{-4},\, 2.0 \times 10^{-4},\, 2.5 \times 10^{-4}$ and $3.0 \times 10^{-4}$
correspond to the weakly chaotic, monotonic $q$ runs in Figs.~\ref{fig_CL_count_1_5},
\ref{fig_CL_count_2_0}, \ref{fig_CL_count_2_5} and ~\ref{fig_CL_count_3_0} respectively. 
The result labeled ``Monotonic" corresponds to the fully chaotic, monotonic $q$ case in Fig.~\ref{fig_CL_count_twist}. 
The two reversed shear cases, ``Reversed 1" and ``Reversed 2", correspond to the fully chaotic, reversed shear cases  in Figs.~\ref{fig_CL_count_nontwist} and ~\ref{fig_CL_count_nontwist_2} respectively.
The steepest dashed line follows the sub-diffusive  scaling in Eq.~(\ref{moments_sca}). The less steep dashed line, corresponds to the scaling $\langle T \rangle_{\max} \sim t^{-1/16}$.}
\label{fig_decay}
\end{figure}

Figure~\ref{fig_count_self_sim} compares the spatio-temporal evolution of the self-similar model in 
Eq.~(\ref{self_sim_T}) in the case of sub-diffusive and diffusive scaling.
In the sub-diffusive case, shown in panel (a), $\gamma=1/2$ and the scaling function is the stretched exponential in Eq.~(\ref{stretched_exp}) with $\nu \approx 1.6$ and $\mu \approx 0.0095$,. In the diffusive case, shown in panel (b), $\gamma=1$ and the scaling function is a Gaussian,
$L_2=A \exp(-\eta^2/\mu)$, with $\mu=0.008$. 
As expected, the transport of the pulse in the diffusive case is significantly faster. 
Note that the iso-contours shown are for the same values of those in Figs.~\ref{fig_CL_count_1_5} indicating good order of magnitude agreement between the sub-diffusive self-similar model and the numerical result in the case of fully chaotic fields with monotonic $q$-profiles. 
As Fig.~\ref{fig_decay} shows, the sub-diffusive scaling of the self-similar model in Eq.~(\ref{moments_sca}) 
agrees well  with the scaling of the decay of 
the temperature maximum observed at short times in all the numerical simulations including the weakly chaotic, and the strongly chaotic cases. The transition to the significantly slower decay, $\langle T \rangle \sim t^{-1/16}$, at longer times after $\chi_\parallel t  > 10^5$ is due to the boundaries and, in the case of weakly chaotic fields, also due to the high order magnetic islands. 

Figure~\ref{fig_delays} also shows the delay
according to the self-similar model  in Eq.~(\ref{self_sim_T}). It is observed that in all cases the delay 
is longer than the sub-diffusive delay and approaches this value as the magnetic field stochasticity increases. 
In particular the delay in the fully chaotic monotonic $q$ case is relatively close to the sub-diffusive delay of the self-similar model. For reference,  the figure also shows the diffusive delay which, as expected, is considerable shorter than the delay found in all the cases studied. 
Note that, the delay at the core shown in Fig.~\ref{fig_delays} takes place in the  $\langle T \rangle = t^{-1/16}$ time scale whereas the delay closer to the edge takes place in the  $\langle T \rangle = t^{-1/4}$  time scale. 
The evolution of the normalized negative flux, $-Q$,  is shown in Fig.~\ref{fig_fluxes}. It is observed that, although the time dependence of  $-Q$ is  qualitatively similar 
in the fully chaotic and in the weakly chaotic cases, the time when the flux reaches its maximum value depends strongly on $\epsilon$. Moreover, the maximum value of the flux, $q_{max}=3.8 \times 10^{-7}$, is orders of magnitudes larger than the values observed in the weakly chaotic cases. 

\subsection{Strongly chaotic fields in reversed shear configurations}

The goal of this subsection is to study the transport properties of heat pulses in fully chaotic magnetic fields in reversed shear configurations. 
Before considering the solution of the parallel heat transport equation for these configurations, we briefly discuss  three unique properties of resonances in reversed shear magnetic fields that will be helpful for the interpretation of the results.  The properties of interest are: separatrix reconnection,  the dependence of the resonances' width with the shear, and the scaling with the amplitude of the perturbation of the  width of resonances in the reversed shear region. 

\begin{figure}
\includegraphics[width=0.45\columnwidth]{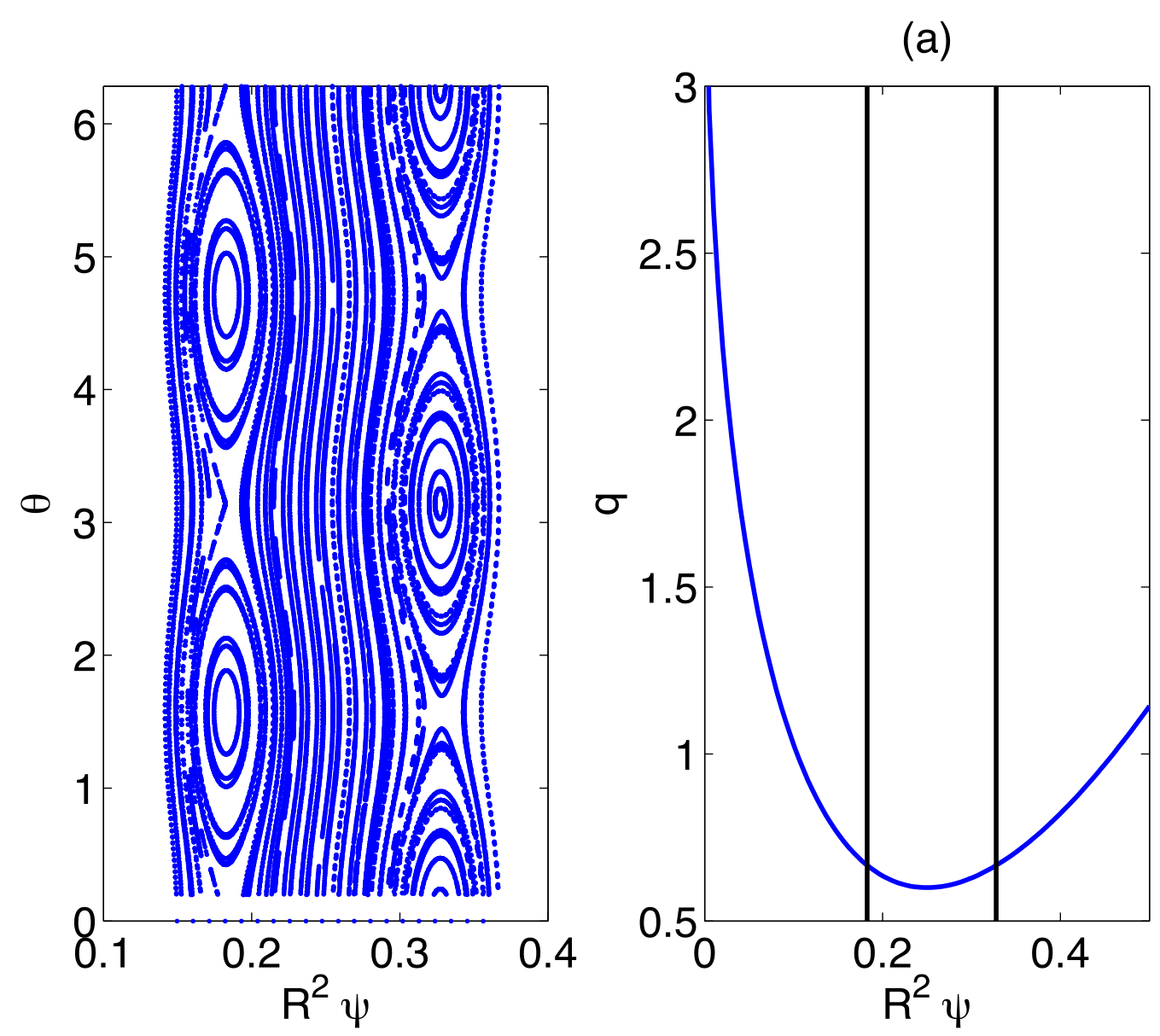}
\includegraphics[width=0.45\columnwidth]{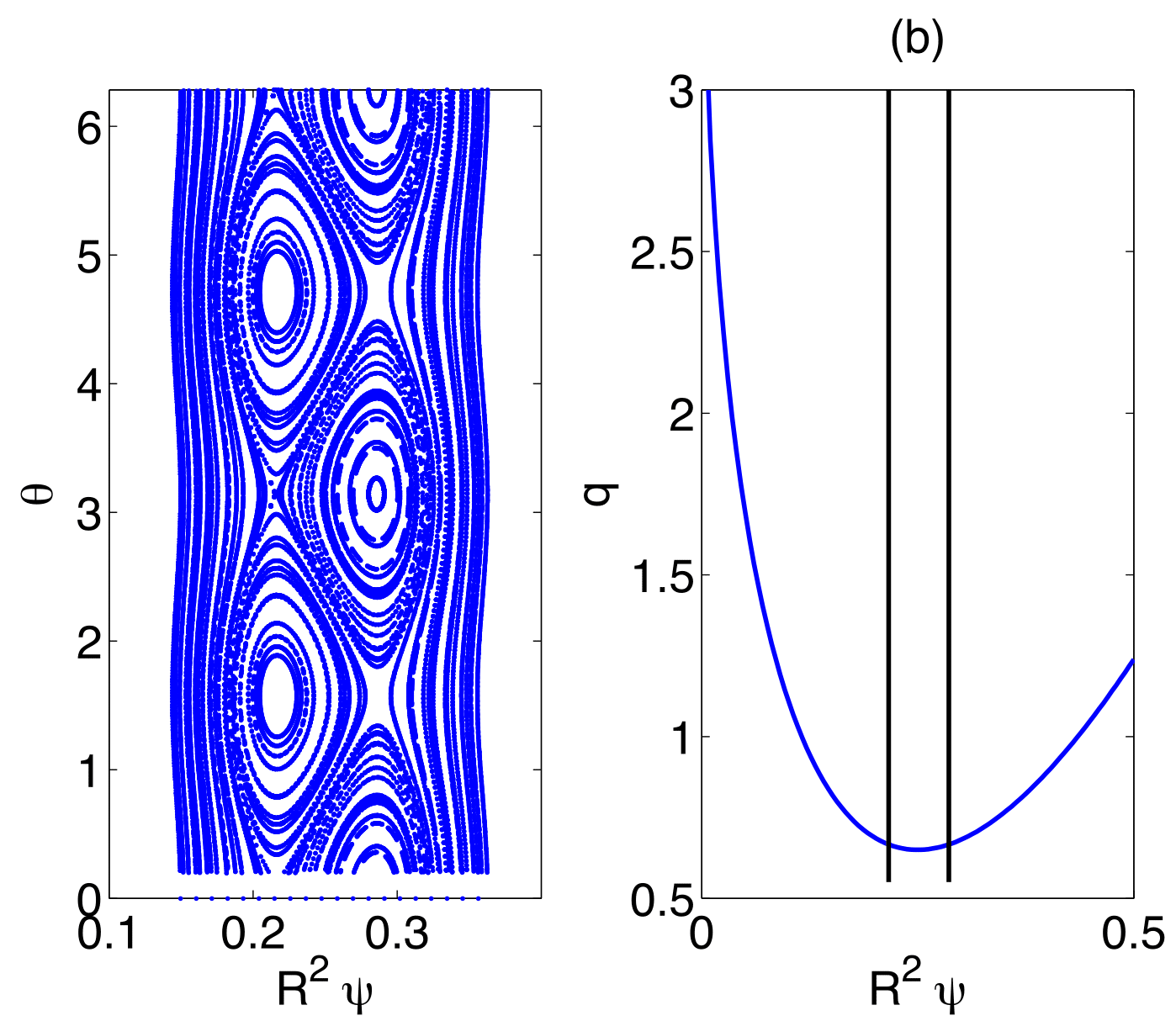}
\includegraphics[width=0.45\columnwidth]{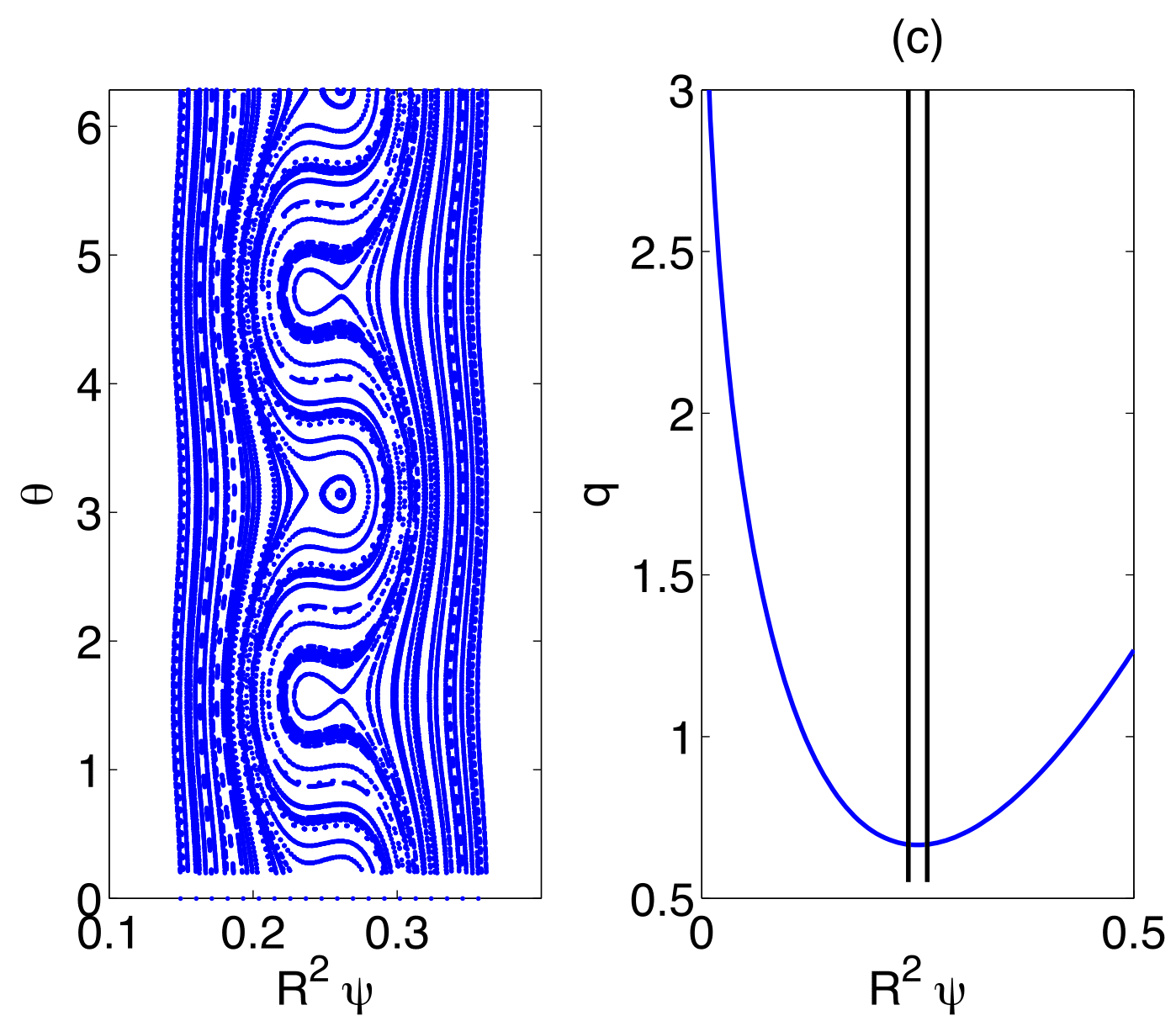}
\includegraphics[width=0.45\columnwidth]{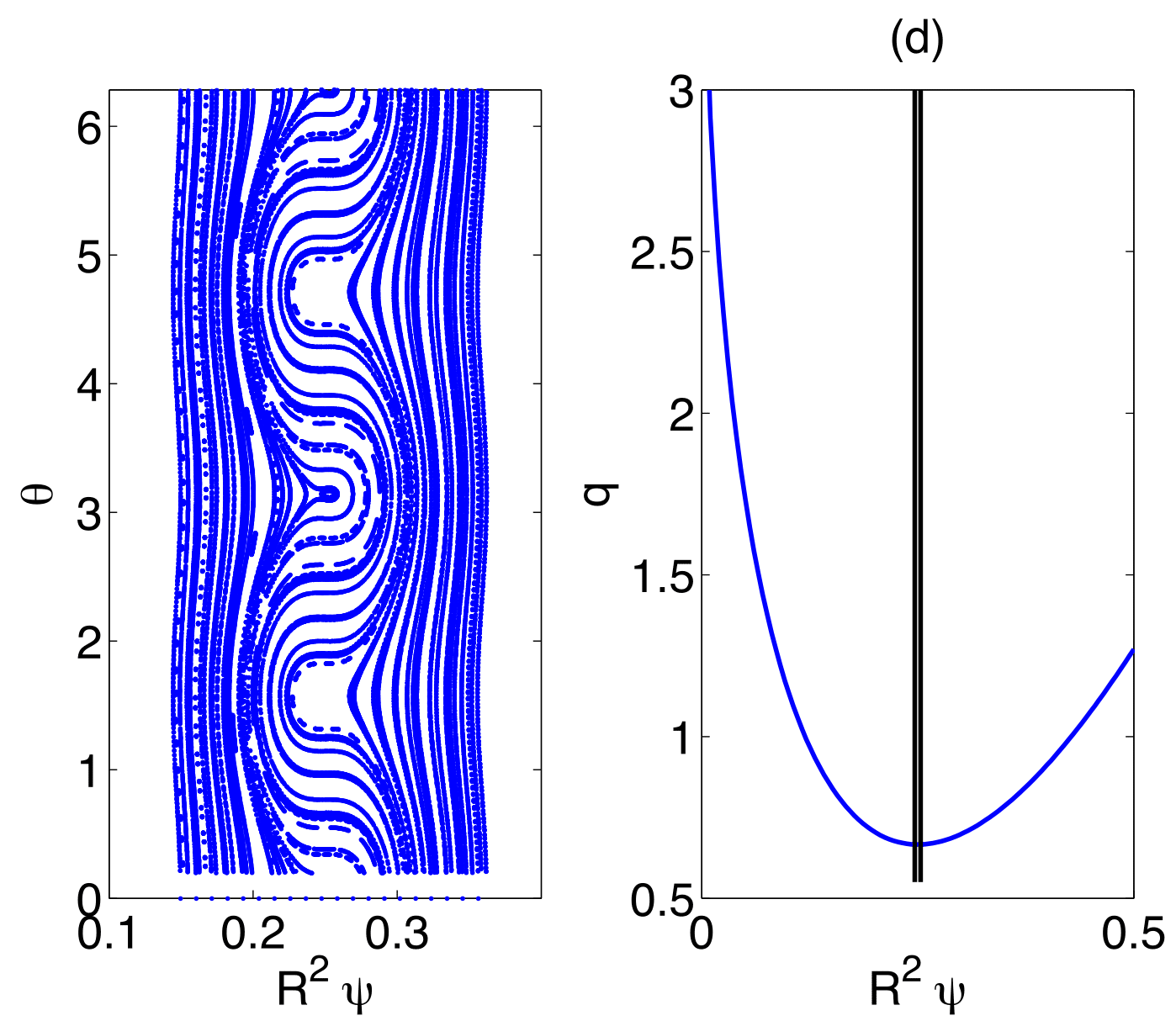}
\caption{(Color online) 
Separatrix reconnection and 
dependence of resonance width on local magnetic shear in reversed shear configurations. 
Each of the four panels shows a Poincare plot of the $(m,n)=(2,3)$ mode, and the corresponding $q$-profile indicating the location of the resonances. In all cases $\epsilon=10^{-4}$ and $\lambda=3.25$. The only parameter varied is $q_0$ in Eq.~(\ref{q_profile}). In (a) $q_0=0.6$, in (b) $q_0=0.65$, in
(c) $q_0=0.665$,  and in (d) $q_0=0.6665$. 
}
\label{qwidth}
\end{figure}

\begin{figure}
\includegraphics[width=0.45\columnwidth]{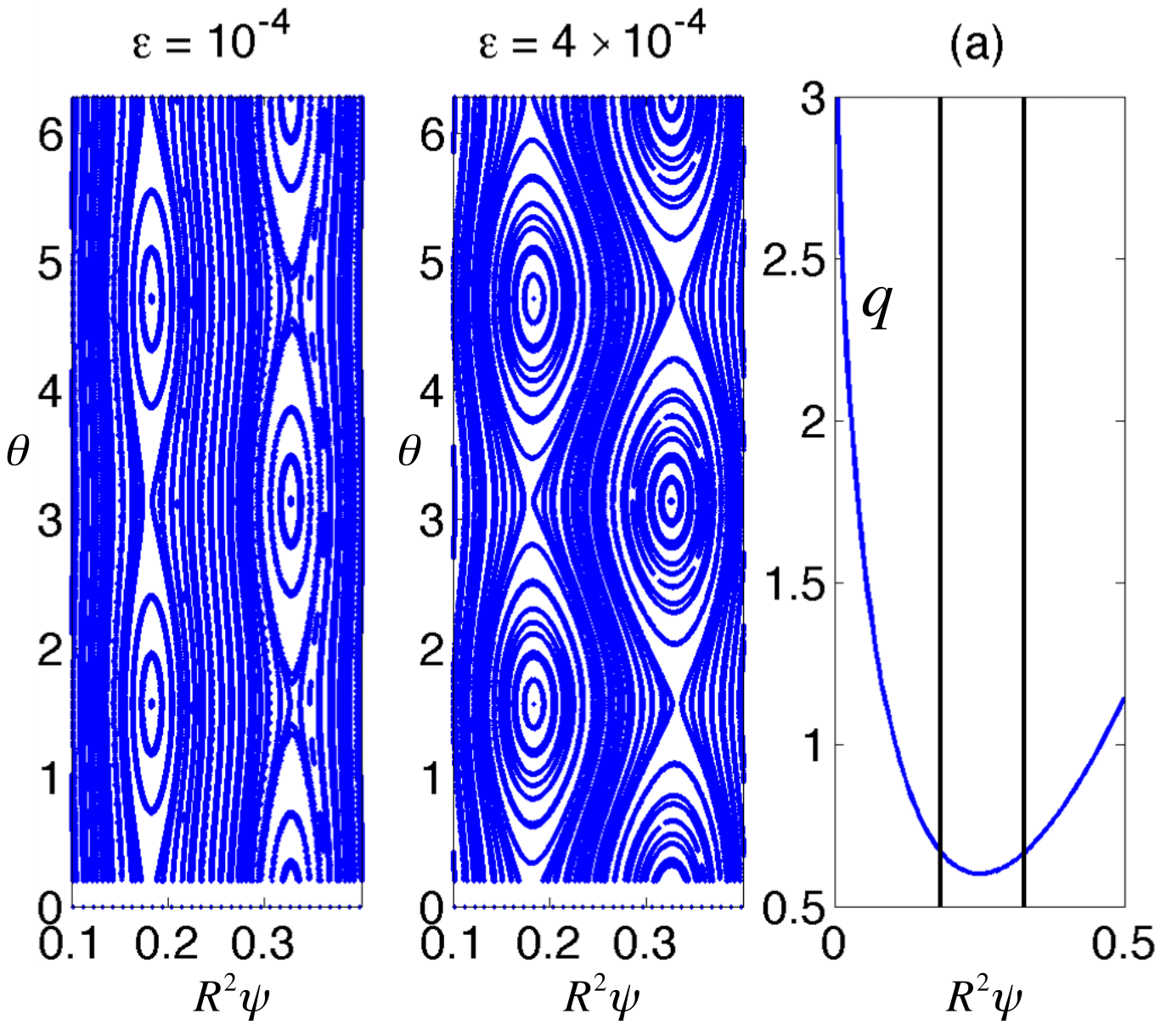}
\includegraphics[width=0.45\columnwidth]{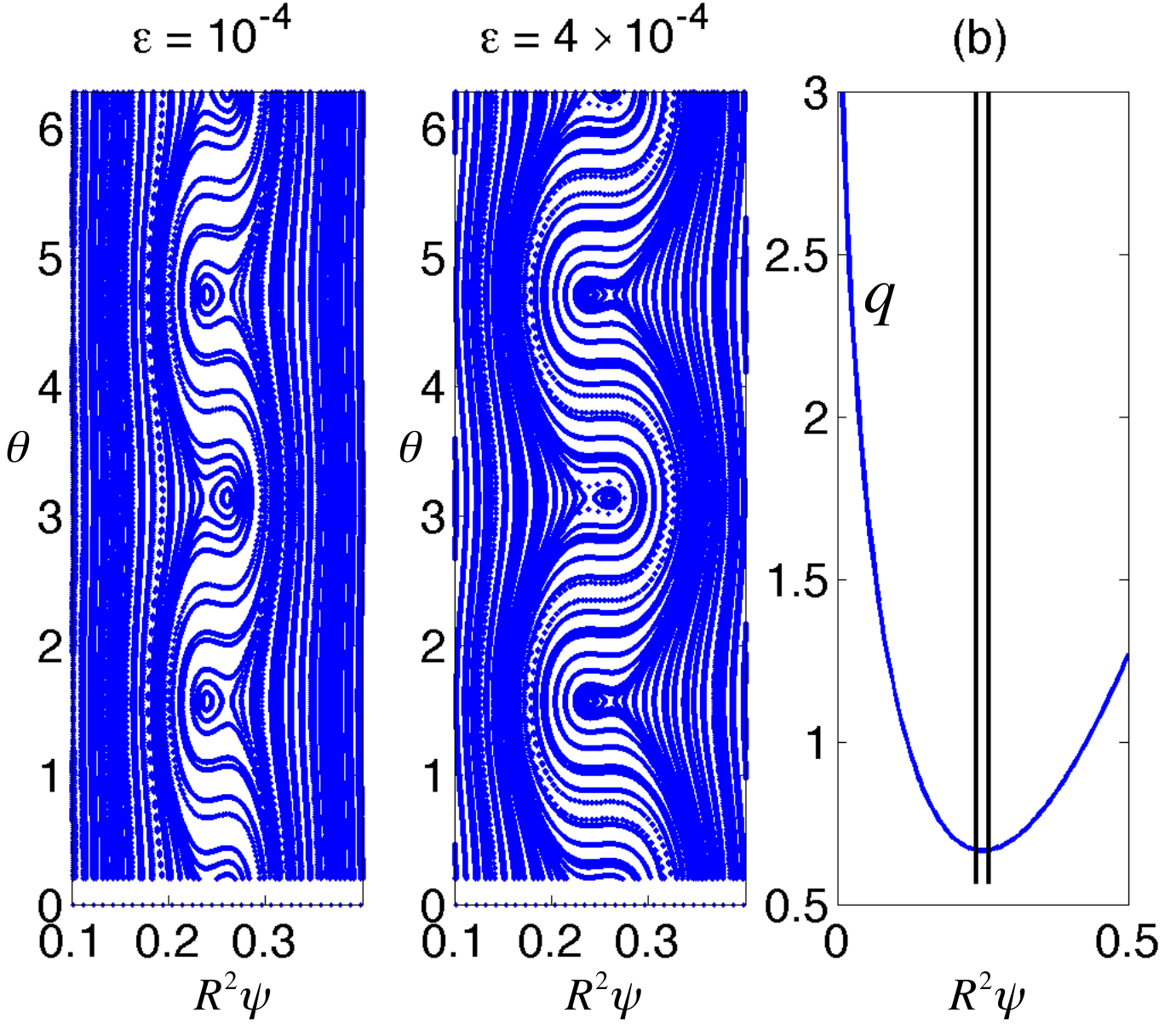}
\caption{(Color online) 
Dependence of resonance's width on magnetic perturbation amplitude, $\epsilon$, in the reversed magnetic shear configuration with $q$-profile in Eq.~(\ref{q_profile}).  
Each of the two panels (a) and (b)  shows Poincare plots of the $(m,n)=(2,3)$ mode for two values of $\epsilon$, along with the corresponding $q$ profile indicating the location of the resonances. 
In panel (a) $q_0=0.6$ and in (b) $q_0=0.665$. In both cases, $\lambda=3.25$. 
As shown in (a), in the finite shear region the  width of the resonances exhibits the usual $\sim \sqrt{\epsilon}$ dependence. However, as shown in (b), near the reversed shear region, this scaling breaks down and the width has a very weak dependence on the amplitude.  
}
\label{qwidth_epsilon}
\end{figure}

Separatrix reconnection is a global bifurcation that changes the topology of twin resonances in the vicinity of the reversed shear region. As shown in Fig.~\ref{qwidth}, because of the non-monotonicity of the $q$-profile, each mode of the magnetic field perturbation creates in general  two, twin magnetic islands with staggered elliptic, o-points.  
In this figure the equilibrium magnetic field had the $q$-profile in 
Eq.~(\ref{q_profile}), and the perturbation consisted of a single mode with $(m,n)=(2,3)$. 
When the twin resonances form relatively far from the reversed shear region, they exhibit the standard heteroclinic topology in  Fig.~\ref{qwidth}-(a). 
However, when the twin resonances approach each other, their stable and unstable manifolds reconnect across the reversed shear region as illustrated in Fig.~\ref{qwidth}-(b).  
When the resonances get closer, the separatrices reconnect once more but in a different way that leads to the homoclinic topology of the twin resonances observed in Fig.~\ref{qwidth}-(c). 
In the Poincare plots in Fig.~\ref{qwidth}, the displacement of the resonances is achieved by changing the value of $q_0$ keeping all the other parameters fixed.
In addition to separatrix reconnection,  
the width of the resonances, $W$, exhibits an interesting dependence on the local magnetic shear, ${\cal S}=d q/ d\psi$. In particular, as can be observed by comparing 
Fig.~\ref{qwidth}-(a) and (b),  before reconnection, $W$ increases when 
 ${\cal S}$ decreases, i.e. as the resonances get closer. However, after reconnection the opposite is the case and, as shown in 
 Figs.~\ref{qwidth}-(c) and (d), $W$ shrinks as ${\cal S}$ decreases. 
 The other property of twin resonances in reversed shear that we want to highlight is
 the dependence of $W$ on the amplitude of the perturbation $\epsilon$. 
 As shown in Fig.~\ref{qwidth_epsilon}-(a), when the resonances 
are far from the reversed shear region and have the heteroclinic topology, they exhibit the usual scaling $W \sim \sqrt{\epsilon}$ . However, as 
shown in Fig.~\ref{qwidth_epsilon}-(b) this scaling breaks down when the resonances are close to the reversed sear region and have the homoclinic topology. In fact, in this case, the size of the trapping region of the resonances is quite small and practically insensitive  to $\epsilon$. 

\begin{figure}
\includegraphics[width=0.55\columnwidth]{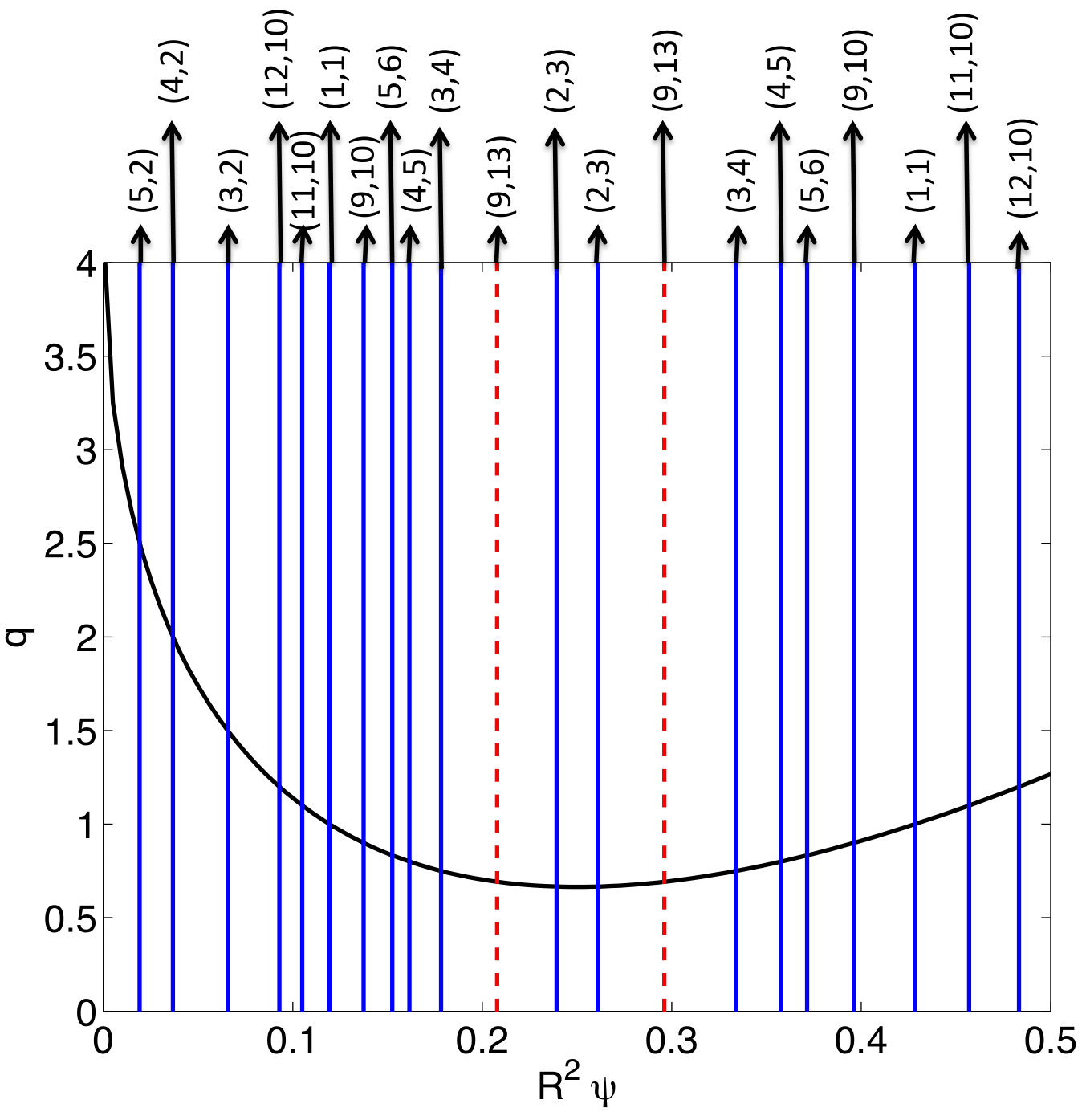}
\caption{
(Color online) $q$-profile and resonant modes used in the numerical study of heat transport in fully chaotic magnetic fields in the reversed shear 
configuration labeled as ``Reversed 1".
The solid line shows the $q$-profile in Eq.~(\ref{q_profile}) as function of $R^2 \psi$.
The modes are given in Eq.~(\ref{modes_nt}), and the vertical lines indicate the corresponding radial location of the resonances.
The red dashed lines correspond to the $(m,n)=(9,13)$ mode. }
\label{qprof_nonmonotonic}
\end{figure}

One of our goals  is to compare the transport properties of heat pulses in reversed shear configurations with the previously discussed results in monotonic 
$q$-profiles, for the case of fully chaotic fields. To make an objective comparison  it is important to guarantee that 
the strength of the perturbation acting in both equlibria are similar. 
The four independent parameters that determine the strength of the perturbation 
in the magnetic field model are: (i) The overall amplitude of the perturbation, $\epsilon$; (ii) The width of the perturbation, $\sigma$; (iii) The number, $N$, of resonances created by the perturbation, and  (iv) The radial distribution of 
the resonances. 
Based on this, we will say that two perturbations have similar strengths when applied to two different $q$-profiles, if they have the same $\epsilon$ and the same $\sigma$, and if they give rise to an equal number of resonances distributed in a similar way in the radial $\psi$ domain
of the two equilibria.  
Following this criterion, here we take $\epsilon = 1 \times 10^{-4}$ and $\sigma=0.5$, which are the same values used in Sec~\ref{strong_twist}. 
For the magnetic field we use the  $q$-profile in 
Eq.~(\ref{q_profile}) with  $q_0=0.665$ and $\lambda=3.25$,
perturbed by the following modes:
\begin{equation} 
\label{modes_nt}\begin{split}
 & (m,n) = \{ (5,2), (4,2), (3,2), (12,10), (11,10),\\
 & (1,1), (9,10), (5,6), (4,5), (3,4), (9,13), (2,3)\} \, .
 \end{split}\end{equation}
Note that we are using less modes than in the monotonic $q$ case in 
Eq.~(\ref{modes}).  This is because in the reversed shear case, typically, 
each mode gives rise to two resonances. In fact, comparing
Figs.~\ref{qprof_nonmonotonic} and  \ref{qprof_monotonic}, it is observed that 
the twelve modes in Eq.~(\ref{modes_nt}) create the same number of resonances in the reversed shear profile in Eq.~(\ref{q_profile}), than the twenty one modes
in Eq.~(\ref{modes}) for the monotonic $q$-profile in
Eq.~(\ref{q_monotonic}). In addition, the spatial distribution of the resonances is similar in both cases, which according to the adopted criterion, implies that the magnetic perturbations acting on the 
 two equilibria have equivalent strengths. 
 %

\begin{figure}
\includegraphics[width=0.53 \columnwidth]{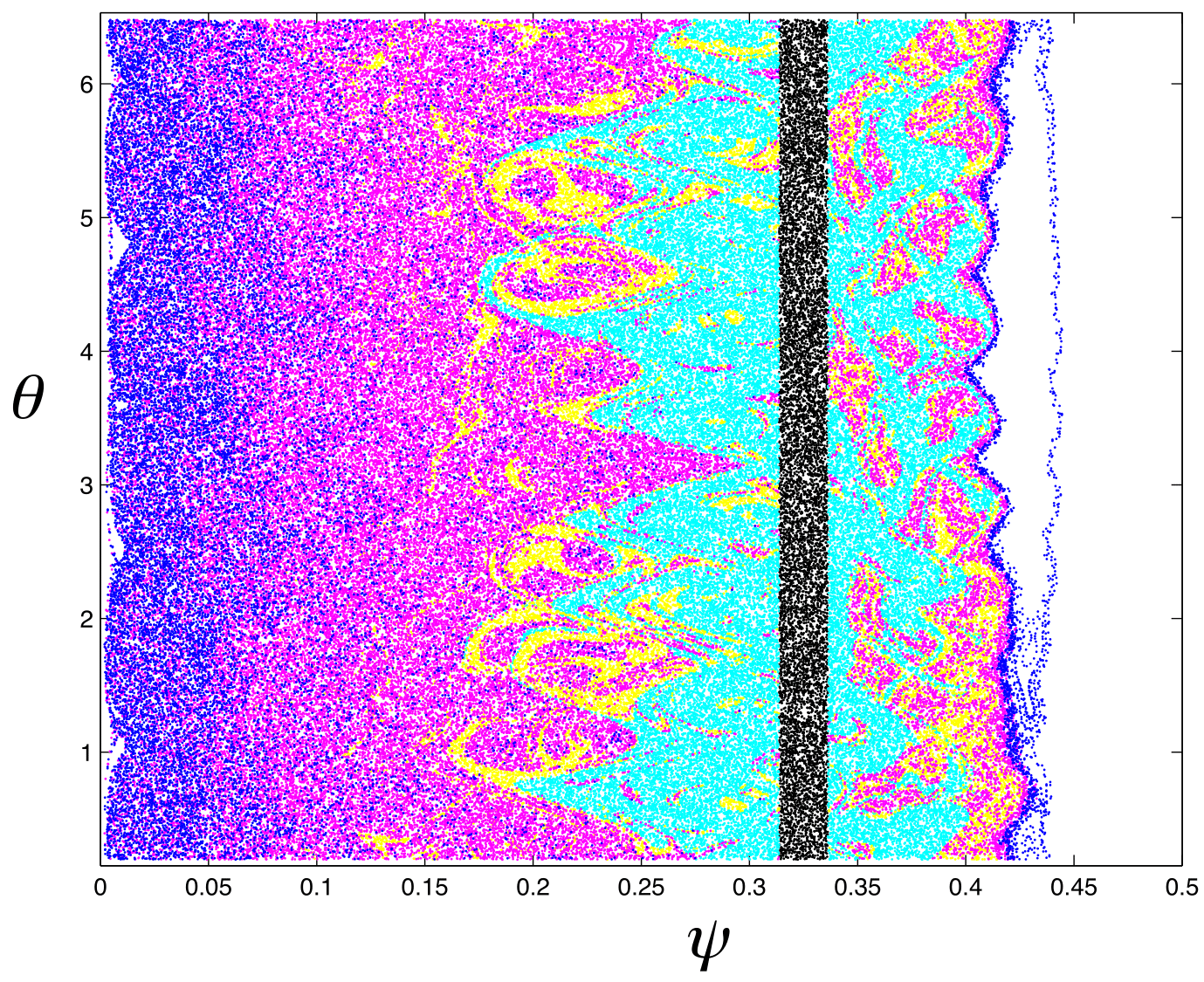}
\includegraphics[width=0.55 \columnwidth]{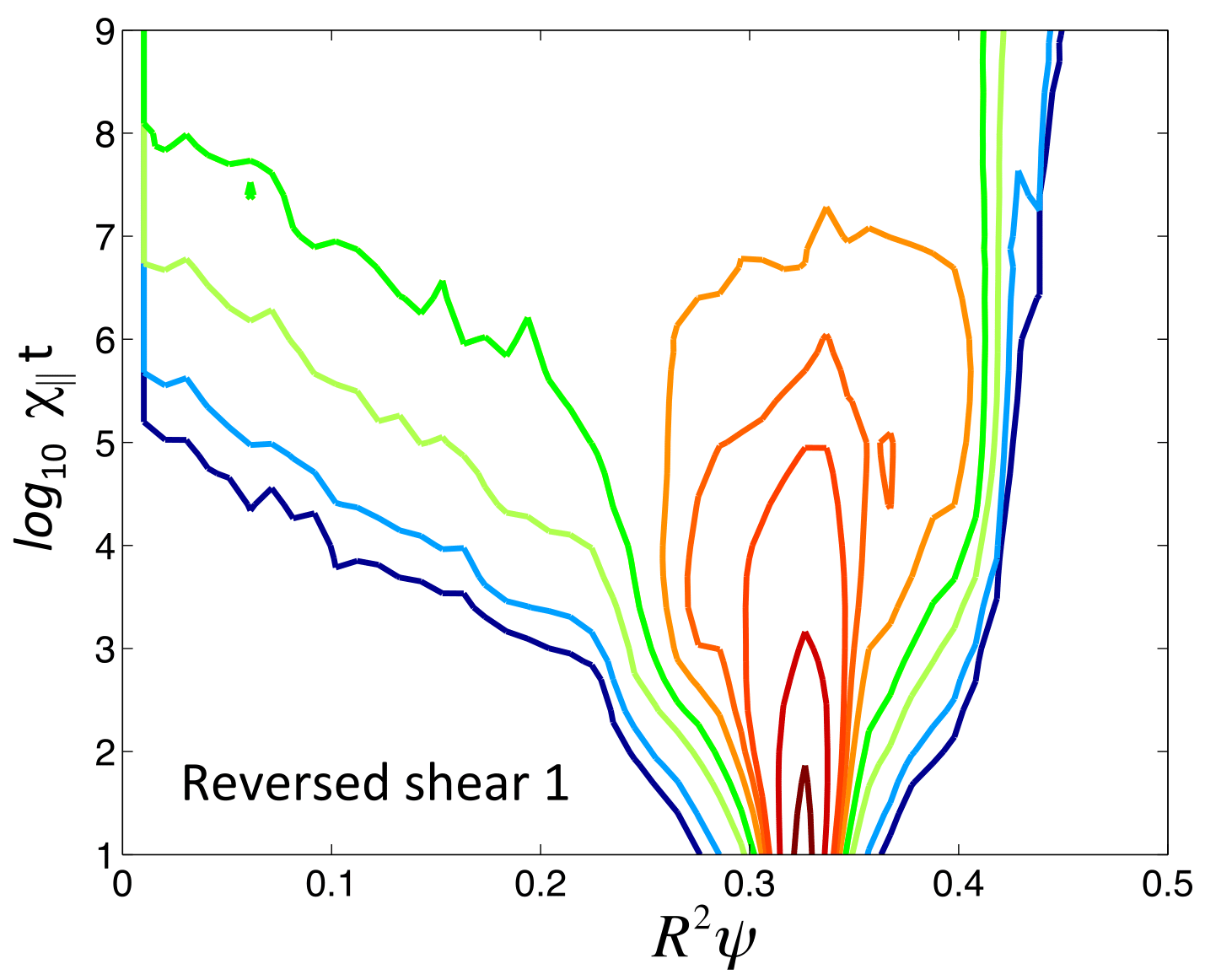}
\caption{
Magnetic field connection length and heat pulse propagation   
in the {\em strongly chaotic} (no flux surfaces) {\em reversed shear} magnetic field with non-monotonic $q$-profile, labeled as ``Reversed 1" in previous figures. 
Top panel shows a Poincare plot  of the magnetic field with $q$-profile in 
Eq.~(\ref{q_profile}) perturbed with the twelve modes in Eq.~(\ref{modes_nt}) with 
amplitude $\epsilon=1\times 10^{-4}$.
As shown in  Fig.~\ref{qprof_nonmonotonic}, due to the nonmonotonicity of the $q$ profile, these twelve modes modes create twenty one resonance
The points in the Poincare plot
correspond to the iterates of a single initial condition, 
 color coded by the value of magnetic field connection length, $\ell_B$, defined in Eq.(\ref{cl}).  The color scale is the same as that used in Figs.~\ref{fig_CL_count_1_5}-\ref{fig_CL_count_3_0}.
The contour plot in the bottom panel shows the spatio-temporal evolution of the heat pulse. Like in Figs.~\ref{fig_CL_count_1_5}-\ref{fig_CL_count_3_0}, the contours correspond to 
(from red to blue) $T_0=\{ 0.5,\, 0.25,\, 0.10,\, 0.075,\,  0.05,\, 0.025,\, 10^{-2},\, 10^{-3},\,  10^{-4} \}$.}
\label{fig_CL_count_nontwist}
\end{figure}

Figure~\ref{fig_CL_count_nontwist} shows a Poincare plot, color coded according to the value of connection length, resulting from the perturbation of the non-monotonic $q$-profile in Eq.~(\ref{q_profile}) by the modes in Eq.~(\ref{modes_nt}). As  expected, the magnetic field is fully chaotic, i.e., there are no flux surfaces nor magnetic islands in the $\psi \in \left(0, 0.425 \right)$ domain. However, the spatial dependence of the connection length, $\ell_B$, is different to the one of the monotonic $q$-profile shown in  Fig.~\ref{fig_CL_count_twist}. 
Overall, the reversed shear connection length is larger than the connection length of the monotonic $q$-profile case, especially in the 
$\psi \sim 0$ neighborhood. This can also be observed  in the radial profiles of the average connection length in Fig.~\ref{fig_CL_vs_psi}, where the reversed shear results are labeled as ``Reversed 1". 
The difference in the connection lengths reflects in the speed and penetration depth of the heat pulse. 
In particular, in the non-monotonic $q$ case the heat pulse slows down in the reversed shear region. This can be appreciated by the stepping in the $R^2 \psi \in (0.2,0.3)$ region  of the temperature iso-countours shown in the bottom panel of Fig.~\ref{fig_CL_count_nontwist}. As a result, as shown in Fig.~\ref{fig_delays}, the time delay of the temperature response is approximately one order of magnitude longer in the reversed shear case  labeled as ``Reversed 1". There are also differences in the flux in Fig.~(\ref{fig_fluxes}).  For example, the peak maximum value of the  flux in the reversed shear case, $q_{max}=3.5 \times 10^{-8}$, is 
ten times smaller than the peak maximum value in the monotonic $q$ case, $q_{max}=3.8 \times 10^{-7}$. Thus, although the strength of the magnetic perturbation is similar, and despite the fact that in both cases there are no flux surfaces or stability islands, there are important difference in the transport properties of heat pulses. In the non-monotonic $q$ case, transport is reduced in the reversed shear region.  
   
The observation that for the same value of $\epsilon$ and $\sigma$, an equal number of similarly distributed resonances leads to less transport in the reversed shear case can be understood in part from the property that, as shown in Figs.~\ref{qwidth} and \ref{qwidth_epsilon}, the width of resonances in  a non-monotonic $q$-profile tends to zero, and has a very weak dependence on $\epsilon$, in the shear reversal region. As a result, resonances in this region exhibit little overlap, chaos is supressed and the flux surfaces are not significantly  affected by the perturbation. 
This behavior is closely related to the typically observed robustness of flux surfaces in non-monotonic $q$-profiles. As discussed in Sec.~\ref{introduction}, at a  fundamental level, this is a generic property of degenerate, non-twist Hamiltonian systems that exhibit very resilient KAM (Kolmogorov-Arnold-Moser) invariant curves (which correspond to magnetic flux surfaces) in the vicinity of reversed shear regions as originally discussed in Refs.~\cite{Dnt0,Dnt1}. 

\begin{figure}
\includegraphics[width=0.45 \columnwidth]{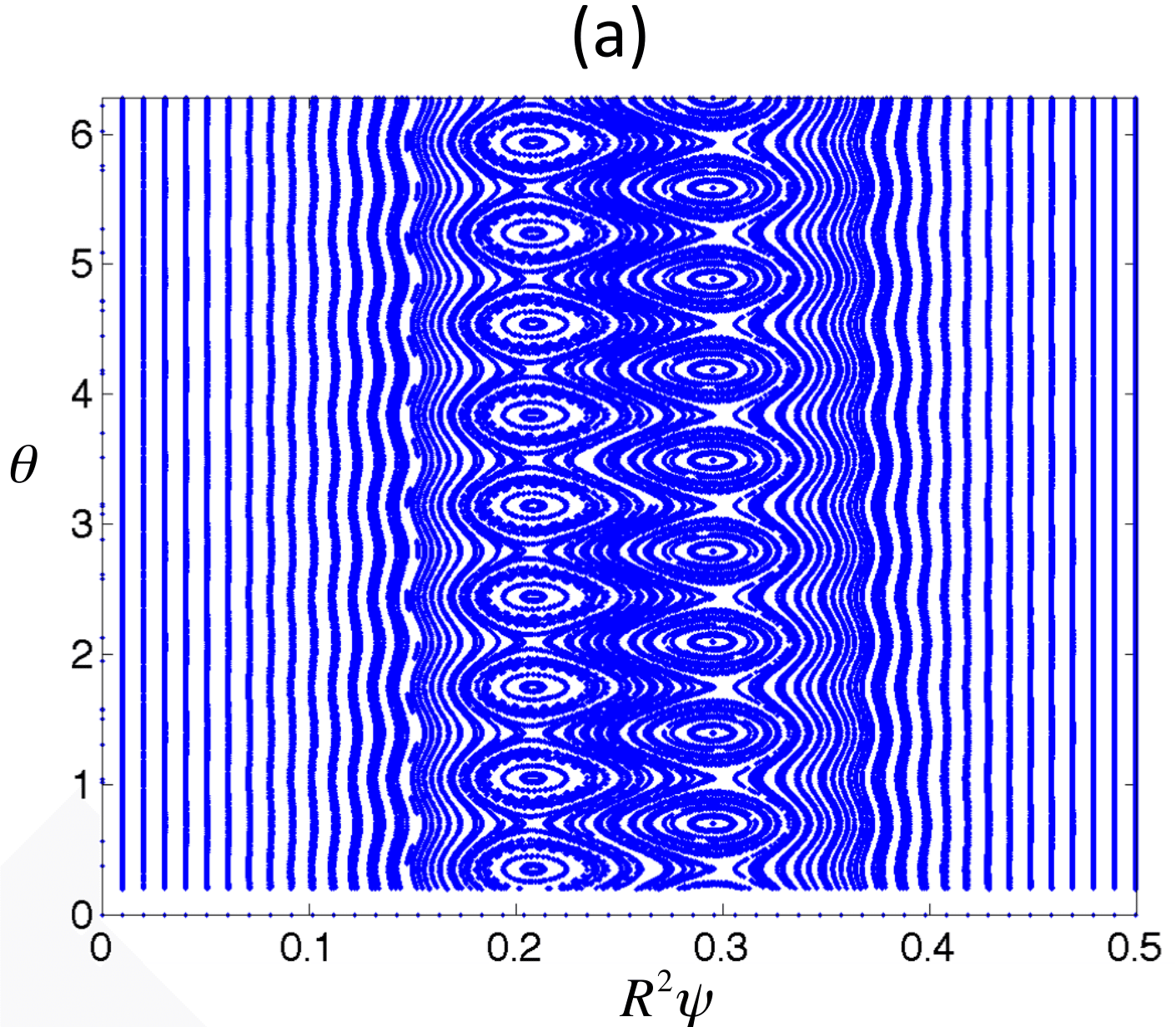}
\includegraphics[width=0.45 \columnwidth]{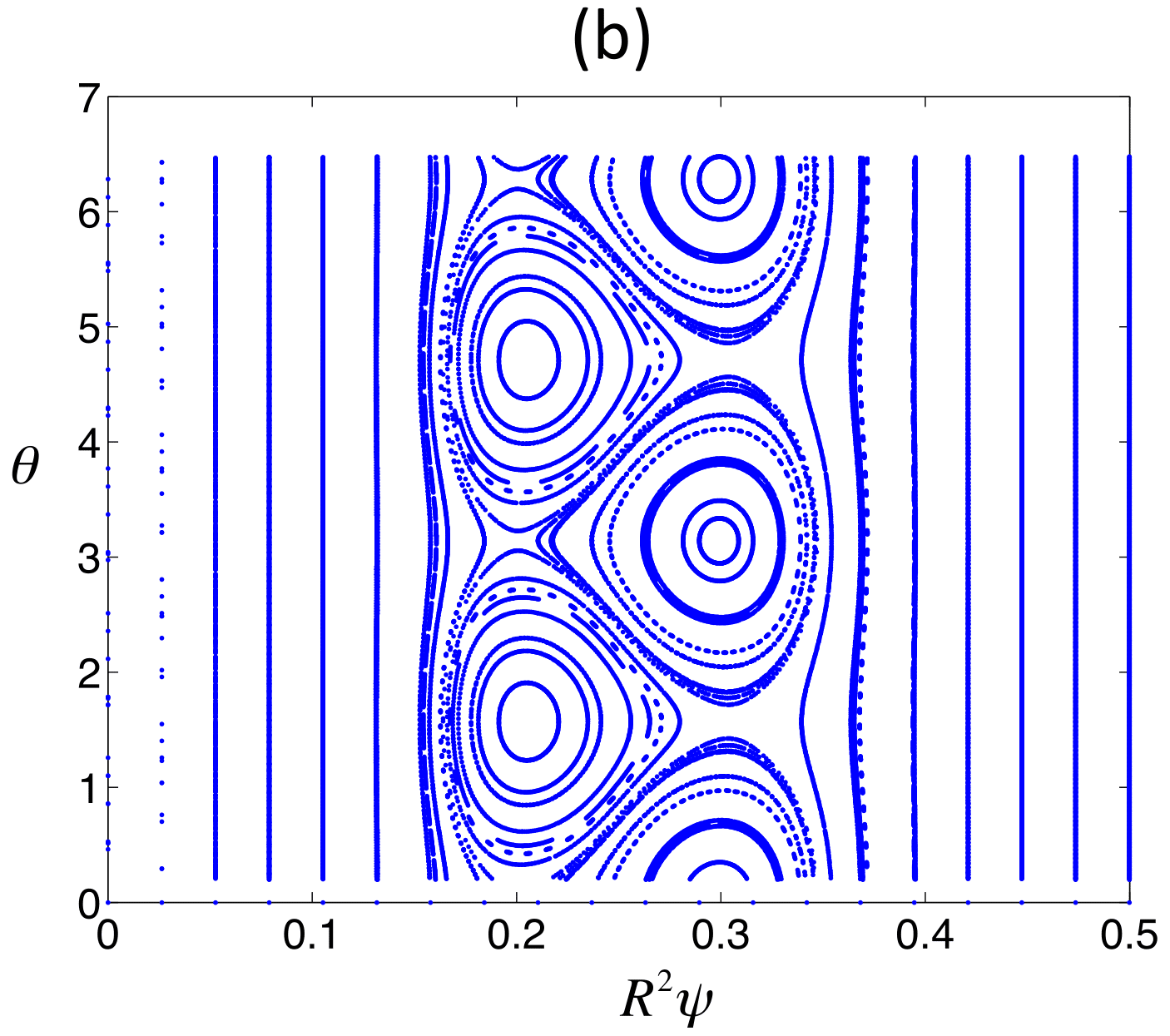}
\caption{
Poincare plots of reconnecting modes in the configurations used in the study of transport in fully chaotic reversed shear magnetic fields. Panel (a) corresponds to the case labeled as ``Reversed 1", and shows the $(m,n)=(9,13)$ mode from the set in Eq.~(\ref{modes_nt}) with 
perturbation parameters
$\epsilon = 1 \times 10^{-4}$ and $\sigma=0.5$, and  non-monotonic $q$-profile in Eq.~(\ref{q_profile})  with $q_0=0.665$,  $\lambda=3.25$.
Panel (b) corresponds to the case labeled as ``Reversed 2", and shows the $(m,n)=(2,3)$ mode from the set in Eq.~(\ref{modes_nt_2}) with 
perturbation parameters
$\epsilon = 4.1 \times 10^{-4}$ and $\sigma=0.05$, and  non-monotonic $q$-profile in Eq.~(\ref{q_profile})  with $q_0=0.64$, and  $\lambda=3.0$.
}
\label{fig_reco_nontwist}
\end{figure}

It is interesting to note that, near the reversed shear region, the distribution of the values of the connection length in Fig.~\ref{fig_CL_count_nontwist} exhibits a kind of ``zig-zag" pattern separating the cyan and magenta colored values of $\ell_B$. The origin of this pattern, which is not observed in the monotonic $q$-case,  can be traced to the existence of separatrix reconnection which, as discussed before, is ubiquitous in non-monotonic $q$-profiles. To verify this intuition, Fig.~\ref{fig_reco_nontwist}-(a) shows the Poincare plot of the $(m,n)=(9,13)$ mode alone. This mode  creates twin resonances in the vicinity of the reversed shear region (see the dashed vertical lines in Fig.~\ref{qprof_nonmonotonic}) which are very close to the reconnection threshold. As observed, the zig-zag pattern follows the $(m,n)=(9,13)$ resonance pattern. 
Note that  (as seen in Fig.~\ref{qprof_nonmonotonic}) in this case there is an inner mode with $(m,n)=(2,3)$. However, this mode appears to have  little effect on the $(m,n)=(9,13)$ resonance and on the level of chaos near the reversed shear region. This can be explained by the 
fact that, as shown in Figs.~\ref{qwidth} and \ref{qwidth_epsilon}, resonances created very close the revered shear region tend to have negligible widths. 

\begin{figure}
\includegraphics[width=0.53 \columnwidth]{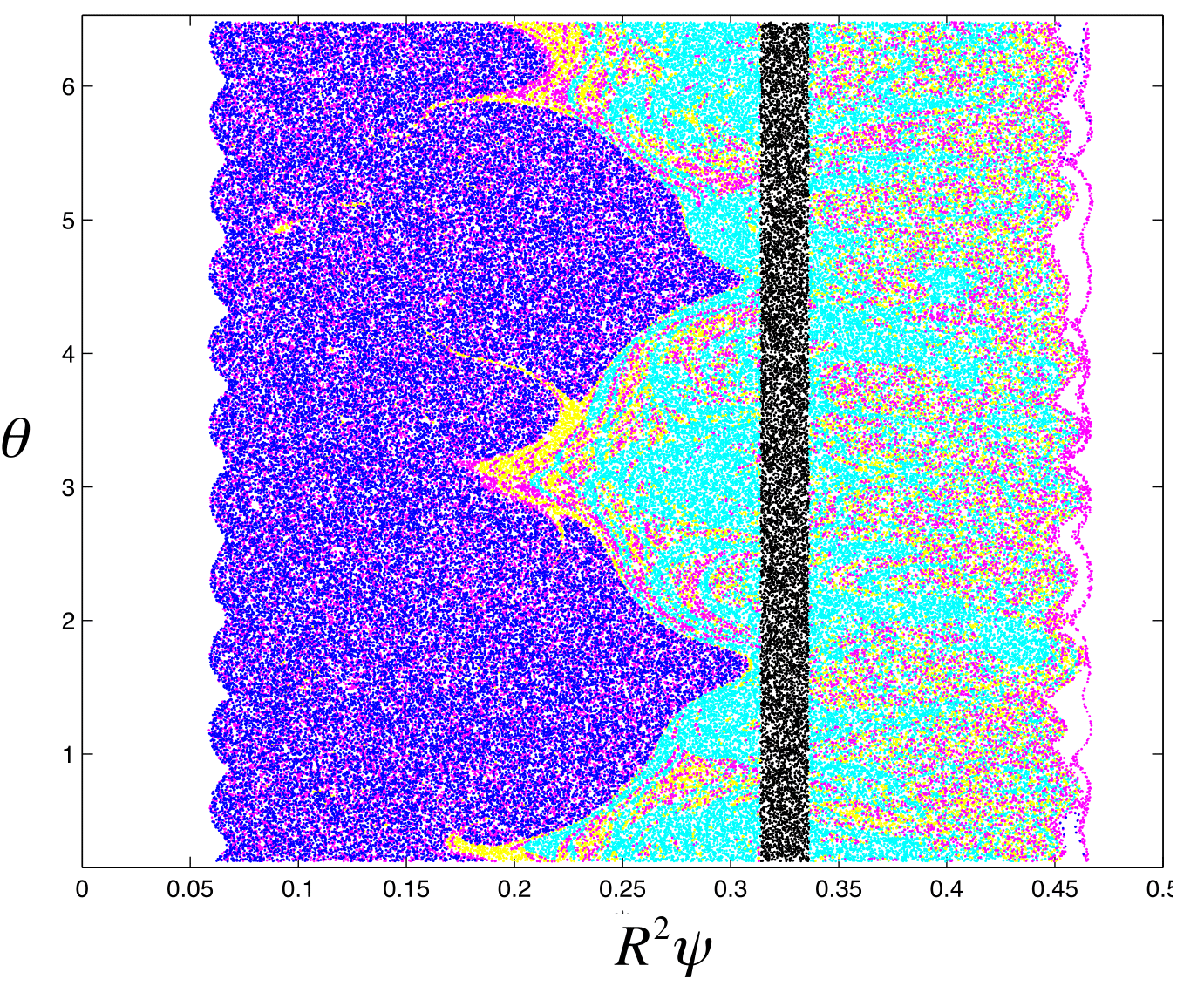}
\includegraphics[width=0.55 \columnwidth]{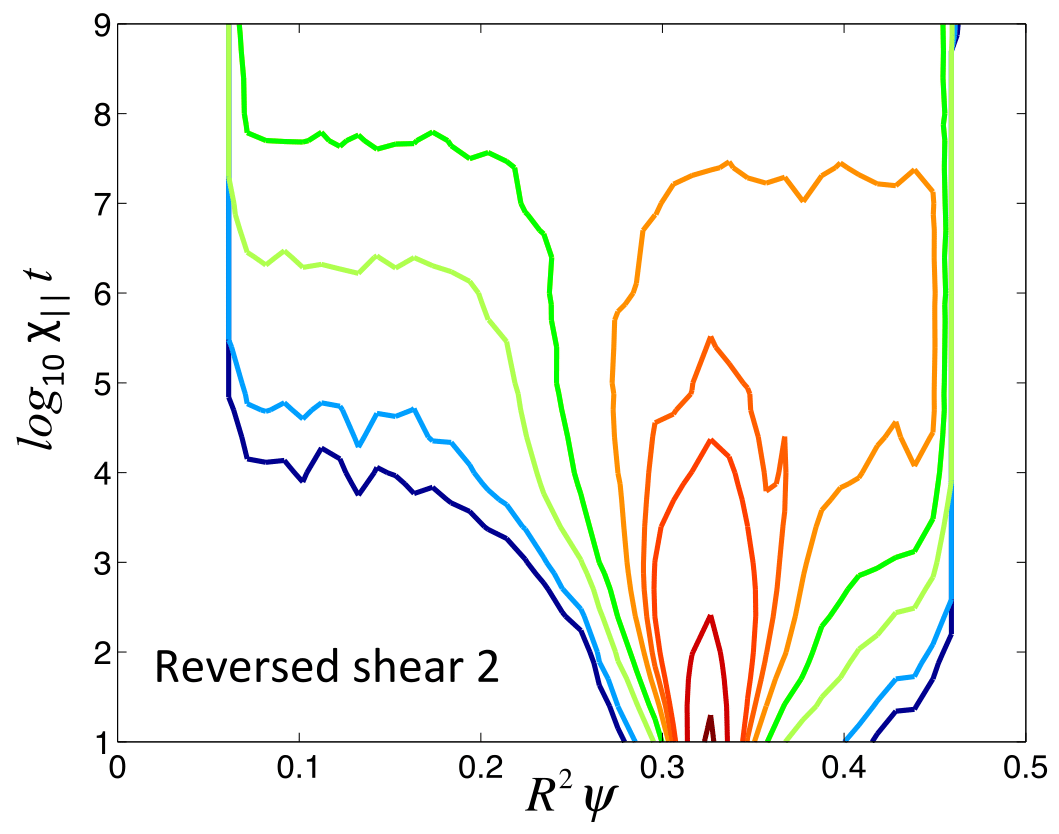}
\caption{
Magnetic field connection length and heat pulse propagation   
in the {\em strongly chaotic} (no flux surfaces) {\em reversed shear} magnetic field with non-monotonic $q$-profile,  labeled as ``Reversed 2" in previous figures. 
Top panel shows a Poincare plot  of the magnetic field with $q$-profile in 
Eq.~(\ref{q_profile}) perturbed by the seventeen modes in Eq.~(\ref{modes_nt_2}) with 
amplitude $\epsilon=4.1\times 10^{-4}$, and $\sigma=0.05$.
The points 
correspond to the iterates of a single initial condition, 
 color coded by the value of magnetic field connection length, $\ell_B$, defined in Eq.(\ref{cl}).  The color scale is the same as that used in Figs.~\ref{fig_CL_count_1_5}-\ref{fig_CL_count_3_0}.
The contour plot in the bottom panel shows the spatio-temporal evolution of the heat pulse. Like in Figs.~\ref{fig_CL_count_1_5}-\ref{fig_CL_count_3_0}, the contours correspond (from red to blue) to 
 $T_0=\{ 0.5,\, 0.25,\, 0.10,\, 0.075,\,  0.05,\, 0.025,\, 10^{-2},\, 10^{-3},\,  10^{-4} \}$.}
\label{fig_CL_count_nontwist_2}
\end{figure}

To further study the role of separatrix reconnection on the connection length and transport, we consider the non-monotonic equilibrium magnetic field in Eq.~(\ref{q_profile}) with $R = 5$, $B_{0}=$1 as before,  but with $q_{0}=$0.64 and $\lambda = 3.0$. The perturbation 
in this case, which is labeled in the figures as ``Reversed 2",  consisted of the following set of modes:
\begin{equation} 
\label{modes_nt_2}\begin{split}
 & (m,n) = \{ (2,3), (7,10), (4,5), (9,10), (13,15), \\
 & (12, 13), (3, 4), (11, 12), (14, 15), (7,8), (8,9), \\
 &(11, 13), (6, 7), (11, 10), (14, 17), (5, 6), (9, 11) \} \, ,
 \end{split}\end{equation}
with $\epsilon=4.1\times 10^{-4}$, and $\sigma=0.05$. 
This case is the same as the one used in Ref.~\cite{dan_diego_2013} to study the role of shearless Cantori on transport. 
Note that the strength of the perturbation is different to the monotonic $q$-profile and ``Reversed 1"  cases considered above as the value of  $\sigma$ is smaller,  the value of $\epsilon$ is larger, and there are more resonances with a different radial distribution.  
In this case,  as shown in Fig.~\ref{fig_reco_nontwist}-(b), it is the $(m,n)=(2,3)$ mode the one that is at the reconnection threshold, and, consistent with the previous description, the connection length shown in Fig.~\ref{fig_CL_count_nontwist_2} exhibits a clear zig-zag pattern that follows closely the resonance pattern of this mode. 
As seen in Fig.~\ref{fig_CL_vs_psi}, the 
average connection length is significantly higher than in fully chaotic monotonic and Reversed 1 configurations. In fact, $\langle \ell_B \rangle$ in the Reversed 2 case is comparable to the 
weakly chaotic monotonic, $\epsilon=3 \times 10^{-4}$, case. As expected, the time delay 
in Fig.~\ref{fig_delays} is also significantly higher, specially in the $\psi \in (0.15,0.225)$ region. 
The flux (see Fig.~\ref{fig_fluxes}) is also remarkably changed. In the Reversed 2 case, $q_{max}=8.6 \times 10^{-9}$, is forty times smaller than the one in the fully chaotic 
monotonic case, $q_{max}=3.8 \times 10^{-7}$. 
As discussed in Ref.~\cite{dan_diego_2013}, the transport properties of the 
Reversed 2 case can be attributed to the presence of  shearless Cantori. These are partial transport
barriers that form near the reversed shear region and significantly slow down transport even in the absence of flux surfaces or stability islands. It is because of these partial barriers that the temperature iso-contours in Fig.~\ref{fig_CL_count_nontwist_2} exhibit a sharp gradient in the $ R^2 \psi \in (0.2, 0.3)$ region. This region also plays a key role in the anomalously slow relaxation of temperature 
gradients \cite{dan_diego_2013}.  To conclude, we mention that, although there is an overall shift, the scaling of the decay rate of the temperature 
maximum, $\langle T \rangle_{max}$, in Fig.~\ref{fig_decay} is quite similar in all cases. 

\section{Summary and conclusions}
\label{conclusions}

We have presented a study of heat pulse propagation in $3$-dimensional chaotic magnetic fields. The study focused on the solution of the parallel heat transport equation. The numerical method was based on the Lagrangian-Green's  function (LG) method that provides an efficient and  accurate technique that circumvents known limitations of  finite elements and finite difference methods. In particular, by construction, the LG method preserves the positivity of the temperature evolution and avoids completely the pollution issues encountered in grid-based methods. 

The magnetic field model consisted of the superposition of an helical field perturbed by resonant modes in cylindrical geometry.
The main two problems addressed were: (i) The dependence of the radial transport of heat pulses on the level of magnetic field stochasticity; and (ii) The role of reversed shear magnetic field configurations on heat transport. 
To addressed the first problem we considered a helical magnetic field with monotonic $q$-profile perturbed by four resonant modes with amplitude $\epsilon$, and solved the corresponding parallel heat transport equation for increasing values of $\epsilon$. 
For the second problem we compared the propagation of heat pulses in monotonic $q$-profile and reversed magnetic shear configurations in fully stochastic fields. 

In the study of the dependence of heat transport on the level of magnetic field stochasticity we considered four different mode amplitudes,
$\epsilon=1.5 \times 10^{-4}$,  $\epsilon=2.0 \times 10^{-4}$, $\epsilon=2.5 \times 10^{-4}$, and $\epsilon=3.0 \times 10^{-4}$.
In all these cases there are no flux surfaces. That is, as shown in the Poincare plots, a single initial condition covers the $R^2 \psi \in (0.05,0.45)$ radial domain. However, the radial transport of the heat pulse is observed to depend strongly on $\epsilon$ due to the  presence of magnetic islands created by high order resonances. In particular, the resonances and Cantori around $R^2 \psi \sim 0.25$ act as quasi-transport barriers, which, in the case 
$\epsilon=1.5 \times 10^{-4}$,  actually preclude the radial penetration of the pulse for the range of time considered. Even for the relatively larger perturbation case 
$\epsilon=3.0 \times 10^{-4}$, the temperature iso-contours exhibit a qualitative change across the $R^2 \psi \sim 0.25$ region. On the other hand, the period-5 islands located around $R^2 \psi \sim 0.15$ do not seem to play a significant role on the radial transport. 
As expected, in the strongly stochastic case corresponding to a monotonic 
$q$-profile perturbed by twenty one strongly overlapping modes  all the high order resonances are practically gone and the radial pulse exhibits an homogeneous spreading towards the core.  

To explore in further detail the role of magnetic field stochasticity on transport, we studied the dependence of the magnetic field connection length, $\ell_B$, on $\epsilon$.  The numerical results showed a very strong, approximately exponential, dependence of $\ell_B$ in the core on the value of $\epsilon$.  There is a  close connection between $\ell_B$ and the transport of localized perturbations in the LG method. In particular, the temperature response at a point ${\bf r}_0$ is a monotonically increasing function of the value of  $\ell_B$ at ${\bf r}_0$. 
As a result, the regions in the $(R^2\psi,\theta)$ plane where $\ell_B$ is large correlate with the regions where the  radial propagation of the heat pulse slows down or stops. In particular, the region 
$R^2 \psi \sim 0.25$ where
the temperature iso-contours exhibit a qualitative change (from almost vertical to almost horizontal) correspond to the region where the average connection length,
$\langle \ell_B \rangle$, exhibits its sharpest gradient. 

In addition to the connection length, we studied the decay rate of the temperature maximum,  $\langle T \rangle_{max}(t)$, the delay of the temperature response, $\tau$, and the 
radial heat flux $\langle {{\bf q}\cdot {\hat e}_\psi} \rangle$,  as functions  of the magnetic field stochasticity. 
In all cases it was observed that  the scaling of $\langle T \rangle_{max}$ with $t$ transitions from sub-diffusive, $\langle T \rangle_{max} \sim t^{-1/4}$, at  short times ($\chi_\parallel  t< 10^5$) to a significantly slower, almost flat scaling at longer times
($\chi_\parallel  t > 10^5$). The  time delay of the response, $\tau$, has a strong dependence on $\epsilon$ and it exhibits a sharp gradient in the transition region $R^2 \psi \sim 0.25$. The 
time evolution of the radial flux, $\langle {{\bf q}\cdot {\hat e}_\psi} \rangle$,
also exhibited a strong dependence on $\epsilon$, in both, the time where the flux achieves its minimum and, most importantly, in the 
magnitude of the flux. In particular the relative small change in $\epsilon$ from $\epsilon=1.5 \times 10^{-4}$ to $\epsilon=3.0 \times 10^{-4}$ 
changes the magnitude of radial flux by several orders of magnitude. 

A study was presented comparing heat transport  in fully chaotic fields with monotonic-$q$ equilibria  with heat transport in fully chaotic fields with reversed shear equilibria. 
By fully chaotic we mean that there are no magnetic flux surfaces nor magnetic stability islands.
Even when the two equilibria configurations experience equal-strength magnetic perturbations, there are clear differences in the transport properties. 
The average connection length in the reversed shear case is twice as as long as the 
average connection length in the monotonic $q$ case near $R^2 \psi \sim 0$. It was also observed that the time delay of the temperature response in the reversed shear case
is approximately one order of magnitude longer than the delay in the monotonic $q$ case. 
There are also differences in the fluxes.  For example, the peak maximum value of the  flux in the reversed shear case is ten times smaller than the peak maximum value in the monotonic-$q$ case.
The culprit of these differences seems to be the presence of 
critical magnetic surfaces (shearless Cantori) and the 
reconnection of resonant modes in the shear reversal region. 
In particular, it was observed that the zig-zag pattern, across which the magnetic field connection length 
experiences a sharp gradient, is caused by the reconnection of twin resonances in the 
shear reversal region. 

A generic, potentially far reaching conclusion that follows from the numerical results presented, is that the connection between heat transport barriers and magnetic field stochasticity is not straightforward. In particular, effective, relatively robust barriers to heat transport can exist (within the time scale of  physical relevance to plasmas confinement) in the case of widespread magnetic field stochasticity. A clear example of this is Fig.~\ref{fig_CL_count_1_5} where it is shown that even in the absence of flux surfaces, the heat pulse in weakly chaotic fields does not penetrate beyond $R^2 \psi \sim 0.25$, for time scales up to $\chi_{\parallel} t \sim 10^9$. The case of reversed shear equilibria is even more striking. For example, as Fig.~\ref{fig_CL_count_nontwist_2} shows, 
even in fully stochastic fields 
with no flux surfaces nor magnetic islands, radial heat transport can slow down considerably going through reversed shear regions. Because of this, beyond the relatively simple study of the existence or destruction of magnetic flux surfaces, the study of heat transport requires a careful study of the effect of the subtle geometry and dynamics of chaotic magnetic fields. 
The application of the ideas and methods used in the present paper 
to the computation of the heat load at the divertor plate in the presence of chaotic fields in toroidal confinement devices is an important problem that will be communicated in a future publication. 

 \section{Acknowledgments}
This work was sponsored by the Office of Fusion Energy Sciences 
of the US Department of 
Energy at Oak Ridge National Laboratory, managed by UT-Battelle, LLC, 
for the U.S.Department of Energy under contract DE-AC05-00OR22725.
%

\begin{thebibliography}{99}

\bibitem{evans_2008}
T.~E. Evans, et. al., Nucl. Fusion {\bf 48} 024002 (2008). 

\bibitem{Schmitz_2008}
O. Schmitz, et. al., Plasma Phys. Control. Fusion {\bf 50} 124029 (2008).

\bibitem{DL}
D.~del Castillo-Negrete and L.~Chac\'on,
Phys. Rev. Lett., {\bf 106}, 195004, (2011).

\bibitem{DL_pop}
D.~del Castillo-Negrete and L.~Chacon,
Physics of Plasmas, {\bf 19}, 056112 (2012).

\bibitem{italians_1}
R. Lorenzini, et al., Nature Physics, {\bf 5}, 825 (1995). 

\bibitem{italians_2}
R. Lorenzini, et al.,   Nuclear Fusion, {\bf 52}, 6, 062004 (2012).

\bibitem{Dnt1}
D.~del Castillo-Negrete, J.~M. Greene, and P.~J. Morrison,
Physica D, {\bf 91} 1 (1996).

\bibitem{dcn_1993}
D. del-Castillo-Negrete,  and P.J. Morrison, Phys. Fluids A, {\bf 5}, 948-965, (1993).

\bibitem{Dnt0}
D.~del Castillo-Negrete,  and P.~J. Morrison, ``Magnetic field line stochasticity and reconnection in a non-monotonic q-profile". Bull. Am. Phys. Soc., Serie II, 37:1543, (1992). 
   
 \bibitem{balescu} 
R. Balescu, Phys. Rev. E, {\bf 58}, 3781Ð3792 (1998).

\bibitem{dcn_2000}
D.~del Castillo-Negrete,  Phys. of Plasmas, {\bf 7}, (5), 1702-1711, (2000).

\bibitem{firpo_2011}
M.~-C. Firpo, and D. Constantinescu,
Phys. Plasmas, {\bf 18}, 032506 (2011).

\bibitem{ibere_2011}
C.~G.~L. Martins, R. Egydio de Carvalho, I.~L. Caldas, and M. Roberto,
Physica A, {\bf 390}, 957 (2011). 

\bibitem{dan_diego_2013}
D. Blazevski and D. del-Castillo-Negrete, Phys. Rev. E {\bf 87}, 063106 (2013).

\bibitem{Ida_2013}
K. Ida et. al., 
New Journal of Physics {\bf 15}  013061 (2013).

\bibitem{held_etal_2001}
  E.~D. Held, J.~D. Callen, C.~C. Hegna, and C.~R. Sovinec,
  Phys. Plasmas, {\bf 8}, 1171, (2001).
 
 \bibitem{del_castillo_2006}
D. del-Castillo-Negrete, 
Phys. of Plasmas {\bf 13}, 082308 (2006).
 
\bibitem{del_castillo_2008} D. del-Castillo-Negrete: ÒNon-diffusive transport in fusion plasmas: fractional diffusion approach.Ó  Chapter in the proceedings of the First ITER Summer School: Turbulent Transport in Fusion Plasmas Edited by Sadri Benkadda, AIP Conference Proceedings, Vol.1013 (2008).

\bibitem{luis_jcp_2013}
L. Chacon, D. del-Castillo-Negrete, and C.~D. Hauck,
``An asymptotic-preserving semi-Lagrangian algorithm for the time-dependent
anisotropic heat transport equation."
Submitted to Journal of Computational Physics, (2013). 

\bibitem{cantori}
R.~S. Mackay, et al., Physica D {\bf 13}, 55-81 (1984). 

\bibitem{hudson} 
S.~R. Hudson, Phys. Rev. E. {\bf 74} 056203 (2006). 

\end{thebibliography}


\end{document}